\documentclass[12pt]{colt2026} 

\usepackage{times}
\usepackage{booktabs}
\usepackage{bm}
\usepackage{tikz}
\usetikzlibrary{arrows.meta,positioning,fit}
\title[Foundations of Generative Solvability]{Foundations of Generative Solvability:\\ Common-Success Coverage and Joint Resource Frontiers}

\coltauthor{%
 \Name{Zichen Song}\\
 \addr City University of Hong Kong (Dongguan), Dongguan, Guangdong, China%
}

\newcommand{\eps}{\varepsilon}
\newcommand{\E}{\mathbb{E}}
\newcommand{\Prob}{\mathbb{P}}

\newcommand{\Z}{\mathbb{Z}}
\newcommand{\R}{\mathbb{R}}
\newcommand{\cD}{\mathcal{D}}
\newcommand{\cH}{\mathcal{H}}
\newcommand{\cS}{\mathcal{S}}
\newcommand{\cX}{\mathcal{X}}
\newcommand{\cA}{\mathcal{A}}
\newcommand{\cQ}{\mathcal{Q}}
\newcommand{\Succ}{\mathrm{Succ}}
\newcommand{\Top}{\mathrm{Top}}
\newcommand{\Vor}{V_{\mathrm{or}}}
\newcommand{\KL}{D_{\mathrm{KL}}}
\newcommand{\dH}{d_{\mathrm{H}}}
\newcommand{\Bin}{\mathrm{Bin}}
\newcommand{\Ind}{\mathbf{1}}
\newcommand{\Leq}{L_{=}}

\makeatletter
\renewcommand\ps@jmlrtps{%
  \let\@mkboth\@gobbletwo
  \def\@oddhead{\scriptsize Preprint\hfill November 2026}%
  \let\@evenhead\@oddhead
  \def\@oddfoot{\scriptsize\copyright\space 2026 Z. Song.\hfill}%
  \let\@evenfoot\@oddfoot
}
\makeatother

\begin{document}

\maketitle

\begin{abstract}%
We develop a theory of how finite experience, frozen knowledge, external evidence, computation, candidate generation and verification jointly determine whether a task can be solved. The central object is \emph{common-success coverage}: every resource-bounded program succeeds on a set of worlds, and a learner must separate only those worlds that remain plausible together yet share no addressable successful program. For finite world families and program menus, a reliability exponent $\gamma$ is achievable with $M$ addressable frozen states if and only if $M\ge K_\gamma$, the least number of programs whose success regions cover every world set of multivariate Chernoff information below $\gamma$; thus $\lceil\log_2K_\gamma\rceil$ bits are necessary and sufficient. Structurally different problems instantiate this principle with closed-form coverage quantities: a continuous covering margin $d_*=(Mw-1)/(M-1)$ with exact exponent $d_*^2/(8\sigma^2)$, a discrete covering order defined by covering designs, and a resource-conversion threshold $D_{\cA}(F)\le2\tau\min(g,v+1)$ for programs with increasing differences. In an end-to-end model, a learner sees noisy single primitives and must solve unseen nonlinear combinations with fresh facts under comparison feedback. We determine its exact frontier at sufficient capacity, a critical resolution scale $s_*$ computable before training, a two-point sandwich at every sample size, second-order sample complexity, and frozen codes of $d\pm O(\log d)$ bits. The theorems yield seven laws of solvability, each tied to its assumptions.%
\end{abstract}


\section{Introduction}\label{sec:intro}

A learned system is trained on finite experience and frozen into a bounded state. At deployment it may read external evidence and its own inputs, compute, propose candidates and query a verifier, and must deliver an answer. Learning theory, complexity theory, decision theory and information-based complexity each measure one of these resources. We ask how they act \emph{jointly}: how do finite experience, reusable knowledge, computation, external information, generation and verification determine whether a task can be solved?
We answer seven concrete questions with theorems (Appendix~\ref{app:extunif}): what must be learned (Q1); which worlds may stay indistinguishable (Q2); when frozen knowledge replaces observation (Q3); when computation is useless (Q4); how generation and verification move the boundary (Q5); why deployment budgets change what must be learned (Q6); and the least capacity preserving a solving ability (Q7).

\paragraph{Common-success coverage.} Identification is not the goal. Every program that respects the deployment budgets succeeds on some set of worlds, its \emph{success region}. After training, a set of worlds remains statistically plausible; if one addressable program succeeds throughout that set, nothing more about it needs to be learned. Learning is needed exactly when plausible worlds cannot share a program. Two features make this more than a slogan. Sharing is a higher-order relation: every three of four worlds may share a program while all four cannot, so the success regions form a simplicial complex that is not determined by its edges. And deployment budgets generate the program space, so evidence, generation and verification change which worlds can share a program and hence what training must resolve.

\paragraph{Contributions.}
\begin{enumerate}
\item \textbf{Foundations (Sections~\ref{sec:foundations}--\ref{sec:coverage}).} An operational contract with eight resources, exact list capacities of three verifier interfaces (Lemma~\ref{lem:capacity}), and the \emph{coverage theorem} (Theorem~\ref{thm:coverage}): for finite world families and menus, exponent $\gamma$ is achievable with $M$ addressable frozen states iff $M\ge K_\gamma$, the least number of programs whose success regions cover every set $H$ with Chernoff information $C(H)<\gamma$. So $\lceil\log_2K_\gamma\rceil$ bits are necessary and sufficient; capacity-induced persistent failure, zero-data solvability and randomized converses follow.
\item \textbf{Geometries (Section~\ref{sec:geometry}).} A continuous \emph{covering margin} $d_*=(Mw-1)/(M-1)$ with exact exponent $d_*^2/(8\sigma^2)$ and a closed-form coverage number; a discrete \emph{covering order} $t_*$ with $K_\gamma=C(m,\ell,t)$, a covering-design number.
\item \textbf{Resource conversion (Section~\ref{sec:conversion}).} For monotone programs with increasing differences, a worldwise guarantee exists iff $g\ge1$ and $D_{\cA}(F)\le2\tau\min(g,v+1)$, via a common worst corner; conversion margins are radii of faces of the success complex.
\item \textbf{End-to-end (Section~\ref{sec:endtoend}).} A learner reads noisy single primitives and must solve unseen nonlinear combinations with fresh facts and comparison feedback. We give the exact frontier, a critical scale $s_*$ computable before training, a two-point sandwich at every $n$ via a cyclic-rotation lemma, second-order sample complexity, and $d\pm O(\log d)$-bit frozen codes.
\end{enumerate}
Section~\ref{sec:laws} relates the structural quantities $C(H)$, $K_\gamma$, $d_*$, $t_*$, $D_{\cA}(F)$, $s_*$ and states seven laws of solvability.

\paragraph{Inherited and new.} We inherit the minimal-incompatible-set exponent and common-mass sandwich of \citet{park2026regret}, of which Theorem~\ref{thm:coverage}(d) is the finite-alphabet worst-case special case, together with multivariate Chernoff information \citep{chernoff1952measure,mishra2024antidistinguishability}, types \citep{csiszar2011information}, list testing \citep{asadi2023list}, two-point methods \citep{lecam1973convergence}, the cycle lemma \citep{dvoretzky1947arrangements}, covering designs \citep{gordon1995covering}, increasing differences \citep{topkis1978minimizing} and Helly-type theorems \citep{alon2024helly}. New are the resource-generated menus and frozen addressing that turn statistical difficulty into a capacity requirement, the covering margin and order, the common-worst-corner conversion theorem, and the end-to-end results. Appendix~\ref{app:attribution} attributes every theorem; Appendix~\ref{app:extended} gives supplementary results, and the remaining appendices give all proofs.

\section{Operational foundations}\label{sec:foundations}

A \emph{world} $\theta$ belongs to a known class $\Theta$. Experience consists of $N$ i.i.d.\ documents with law $P_\theta$ on $\cX$. A \emph{training rule} $\mathcal A$ reads at most $T_{\mathrm{read}}$ documents, uses at most $S$ bits of workspace and $T_{\mathrm{ops}}$ word operations, and writes once a \emph{frozen state} $W\in\{0,1\}^B$; it is a function of $X_{1:n}$, $n=\min(N,T_{\mathrm{read}})$, and internal randomness.

A \emph{task} consists of adversarial parameters $Q\in\cQ$ and task randomness $\xi$, independent of $\theta$ and the corpus, which includes the \emph{fresh facts}. A \emph{deployment controller} $\pi$ receives $(W,Q)$ and interleaves, under per-path budgets, evidence reads about $\xi$ (budget $e$), accesses (budget $c$; probes of frozen bits or reads of local input bits, fixed per section), candidate generations ($g$) and verifier queries about generated candidates ($v$).
It must deliver a generated candidate; tolerances satisfy $\tau\ge0$, and deployment workspace and computation are counted only as sufficient budgets. We write $V(\mathcal A,\pi;\theta,Q)$ for the success probability over corpus, task randomness and internal randomness, and $\omega=(e,c,g,v)$ for the online budgets.

\begin{definition}[Frontier, oracle and regret]\label{def:frontier}
The worst-case solvability frontier is
\[
V^*(N,T_{\mathrm{read}},B,S;\,e,c,g,v)=\sup_{(\mathcal A,\pi)}\ \inf_{\theta\in\Theta,\,Q\in\cQ}\ V(\mathcal A,\pi;\theta,Q),
\]
where the supremum ranges over public pairs fixed before $\theta$ and $Q$ and respecting all budgets. The \emph{same-budget oracle} $\Vor(\theta,Q;\omega)$ is the best success probability of a controller that knows $\theta$, has the same online budgets, and knows nothing about $\xi$ beyond what those budgets reveal. The \emph{oracle-relative regret} of a pair is $\sup_{\theta,Q}[\Vor(\theta,Q;\omega)-V(\mathcal A,\pi;\theta,Q)]$.
\end{definition}

Throughout, (C1) all algorithms and interfaces are fixed before $\theta$; (C2) world-dependent information enters only through documents, the frozen state or online observations; (C3) budgets are per path; (C4) the final answer must have been generated; (C5) absolute success and oracle-relative regret are different objects; and (C6) knowing $\theta$ does not reveal the fresh facts. Appendix~\ref{app:extfound} separates five layers of indistinguishability and fixes the units of each budget in each section (Table~\ref{tab:units}).

\paragraph{Feedback structure.} An \emph{equality} verifier answers whether a generated candidate equals the target; a \emph{radius} verifier whether it lies within $\tau$ of a real target; a \emph{comparison} verifier answers $<$, $=$ or $>$ for an integer target.

\begin{lemma}[List capacity of verification]\label{lem:capacity}
Consider a deterministic controller with per-path budgets $(g,v)$ that faces an unknown target, may verify only generated candidates, and must deliver a generated candidate. Put $\nu=\min(g,v+1)$.
\begin{enumerate}
\item[(a)] With an equality verifier, the set of targets on which it succeeds has at most $\Leq:=\nu$ elements if $g\ge1$, and is empty if $g=0$.
\item[(b)] With a radius-$\tau$ verifier on $\R$, the success set is contained in a union of at most $\nu$ closed intervals of length $2\tau$.
\item[(c)] With a comparison verifier on $\Z$, the success set has at most $L:=2^{\nu}-1$ elements.
\end{enumerate}
In each case every set of the stated form is the success set of some controller that uses at most $\nu$ generations and $\nu-1$ verifications on each path. Consequently, against a target with law $\mu$ independent of the controller's randomness, the optimal success probability is $\Top_{\Leq}(\mu)$ in case (a), the largest $\mu$-mass of a union of $\nu$ intervals of length $2\tau$ in case (b), and $\Top_L(\mu)$ in case (c), where $\Top_L(\mu)=\max_{|S|\le L}\mu(S)$.
\end{lemma}

For (a) and (b) the proof (Appendix~\ref{app:capacity}) follows the all-negative feedback path; for (c) it solves $F(g,v)\le\max\{1,1+2F(g-1,v-1)\}$. A comparison splits the remaining candidates into two live subtrees, whereas a ``no'' from an equality or radius verifier removes one candidate. Feedback structure, not the number of returned bits, sets the capacity.

\section{Common-success coverage}\label{sec:coverage}

Fix online budgets $\omega$. A \emph{deployment policy} $\pi$ maps each task to a controller that respects $\omega$, and $\Pi_\omega$ denotes a set of such policies, the \emph{program menu}. A \emph{target} assigns to each pair $(\pi,\theta)$ whether $\pi$ meets the requirement in world $\theta$ for every task. Examples are worldwise exact success (Sections~\ref{sec:geometry} and~\ref{sec:conversion}) and oracle-optimal success (Section~\ref{sec:endtoend}).

\begin{definition}[Coverage objects]\label{def:coverage}
\begin{enumerate}
\item The \emph{success region} of $\pi$ is $\Succ(\pi)=\{\theta:\pi\text{ meets the target in }\theta\}$.
\item A set $H\subseteq\Theta$ is \emph{compatible} if $H\subseteq\Succ(\pi)$ for some $\pi\in\Pi_\omega$, and \emph{incompatible} otherwise. It is \emph{minimally incompatible} if it is incompatible and all of its proper subsets are compatible.
\item The \emph{success complex} $\cS_\omega$ is the family of compatible sets. It is closed under taking subsets, that is, an abstract simplicial complex whose maximal faces are maximal success regions.
\item The \emph{task-relevant equivalence} $\theta\equiv_\omega\theta'$ holds iff $\theta,\theta'$ lie in exactly the same success regions. Pairwise compatibility is in general not transitive and is never called an equivalence.
\item The \emph{common statistical difficulty} of a nonempty finite $H$ is the multivariate Chernoff information (Lemma~\ref{lem:minimax})
\[C(H)=\min_Q\max_{\theta\in H}\KL(Q\|P_\theta).\]
\item For $\gamma>0$, the \emph{difficulty complex} is $\cH_\gamma=\{H\ne\emptyset:\,C(H)<\gamma\}$. It is closed under subsets because $C$ is monotone under inclusion.
\item The \emph{coverage number} $K_\gamma(\omega)$ is the least number of policies in $\Pi_\omega$ whose success regions contain every member of $\cH_\gamma$ ($\infty$ if none). It is nondecreasing in $\gamma$; $K_{0+}=\lim_{\gamma\downarrow0}K_\gamma$ is, for distinct laws, the least number of success regions covering $\Theta$.
\end{enumerate}
\end{definition}

The success complex is generally not the clique complex of its compatible pairs (Proposition~\ref{prop:highorder}).

\paragraph{Addressable states.} A scheme with $M$ \emph{addressable states} is a training rule mapping $X_{1:n}$ and internal randomness to $W\in[M]$ with a public assignment $\phi_n:[M]\to\Pi_\omega$ (deployment may randomize). $B$ frozen bits address at most $2^B$ policies; with a single task and yes/no feedback, $c$ probes address at most $2^{\min(B,c)}$ success sets (Appendix~\ref{app:extcov}). The risk is $R_n=\max_\theta\Prob_\theta(\theta\notin\Succ(\pi_W))$.

\begin{theorem}[Common-success coverage theorem]\label{thm:coverage}
Let $\Theta$ be finite, let the laws $P_\theta$ on a finite alphabet $\cX$ be mutually absolutely continuous, and let the menu $\Pi_\omega$ be finite. Fix $\gamma>0$ and $M\ge1$.
\begin{enumerate}
\item[(a)] \emph{Achievability.} If $K_\gamma\le M$, there is a scheme with $M$ states and deterministic deployment such that $R_n\le(n+1)^{|\cX|}e^{-n\gamma}$ for every $n\ge1$ (for $n=0$ the bound is trivial).
\item[(b)] \emph{Converse.} If $K_\gamma>M$, every scheme with $M$ states satisfies, for every $n\ge0$,
\[
R_n\ \ge\ \frac{\eta_\Theta}{M}\,\min_{H\in\cH_\gamma}\int\min_{\theta\in H}P_\theta^{\otimes n},\qquad\text{where }\ \lim_{n\to\infty}-\tfrac1n\log\int\min_{\theta\in H}P_\theta^{\otimes n}=C(H).
\]
This holds for any training rule and for assignments that vary with $n$, with $\eta_\Theta=1$ for deterministic deployment and $\eta_\Theta=(|\Theta|2^{|\Theta|})^{-1}$ if deployment may randomize over policies.
\item[(c)] \emph{Exponent and capacity.} The exponent $\gamma$ is achievable, in the sense $\liminf_n-\frac1n\log R_n\ge\gamma$, if and only if $K_\gamma\le M$. Vanishing risk $R_n\to0$ is possible if and only if $M\ge K_{0+}:=\lim_{\gamma\downarrow0}K_\gamma$, and then some positive exponent is achievable; if $M<K_{0+}$, then $R_n\ge\eta_\Theta/M$ for every $n$. The optimal exponent with $M$ states is
\[
E^*(M)=\sup\{\gamma>0:\ K_\gamma\le M\}\qquad(\sup\emptyset=0),
\]
and this supremum is attained whenever $0<E^*(M)<\infty$. With fully readable states, $\lceil\log_2K_\gamma\rceil$ frozen bits are necessary and sufficient for exponent $\gamma$.
\item[(d)] \emph{Unlimited memory.} If $M\ge|\Pi_\omega|$, then $E^*(M)=\min\{C(G):\ G\ \text{minimally incompatible}\}$, and $E^*(M)=\infty$ if $\Theta$ is compatible.
\item[(e)] \emph{Zero data and persistent failure.} If $\Theta$ is compatible, then $R_n=0$ for every $n\ge0$ with $B=0$. If some $H$ with $C(H)=0$, that is a set of worlds with identical laws, is incompatible, then every scheme has $R_n\ge1/|H|$ for every $n$, $M$, $S$ and amount of computation.
\end{enumerate}
\end{theorem}

\begin{proof}[Proof outline (Appendix~\ref{app:coverage})]
For (a), training computes the empirical type $\hat Q_n$; the plausible set $\{\theta:\KL(\hat Q_n\|P_\theta)<\gamma\}$ lies in $\cH_\gamma$, so one of $K_\gamma$ chosen regions covers it, and failure has probability at most $(n+1)^{|\cX|}e^{-n\gamma}$. For (b), the $M$ regions miss some $H\in\cH_\gamma$; one witness world of $H$ per state gives the common likelihood mass. Part (c) combines both, (d) uses enough states, and (e) uses that identical laws induce identical laws of $W$.
\end{proof}

\paragraph{Interpretation.} The learner must separate exactly the worlds that lie together in a set of difficulty below $\gamma$ but in no common success region; all other indistinguishability is harmless, and $\lceil\log_2K_\gamma\rceil$ can be far below $\log_2|\Theta|$. Applying \citet{park2026regret} to every $M$-subset of the menu gives the same $E^*(M)$, so (c) is not a new exponent principle; its contribution is the covering formulation, which reduces capacity to one combinatorial number that the structured models compute in closed form, and handles randomized deployment and every $n$. Appendix~\ref{app:extcov} gives oracle-relative regret (Proposition~\ref{prop:oracle}), set-cover hardness of $K_\gamma$ (Proposition~\ref{prop:setcover}), monotonicity in budgets and its failure (Proposition~\ref{prop:monotone}), and the information barrier (Proposition~\ref{prop:barrier}).

\section{Geometries of solvability}\label{sec:geometry}

Theorem~\ref{thm:coverage} does not compute $K_\gamma$ and does not apply to infinite world families. We treat two structurally different problems in which the covering question has a closed-form answer that controls learning at every sample size.

\subsection{Continuous worlds: the covering margin}\label{sec:continuous}

\paragraph{Contract.} A parameter $\theta\in[0,1]$ is observed through $N\ge0$ measurements $Y_i\sim\mathcal N(\theta,\sigma^2)$. The task is to deliver a generated number within $\tau>0$ of $\theta$, with $g\ge1$ generations and $v$ queries to an exact radius-$\tau$ verifier. The frozen state has $B$ bits of which at most $c$ are probed, and deployment is deterministic (randomization changes the constants; Appendix~\ref{app:extgeo}). By the all-``no'' path, deployment distinguishes at most $M=2^{\min(B,c)}$ states. Put
\[
\Leq=\min(g,v+1),\qquad w=2\Leq\tau .
\]
Let $R_N$ be the least worst-case failure probability with $N$ measurements. By Lemma~\ref{lem:capacity}(b), each state succeeds on a union of at most $\Leq$ intervals of total length at most $w$.

\begin{theorem}[Continuous covering margin]\label{thm:continuous}
\begin{enumerate}
\item[(i)] If $w\ge1$, then $R_N=0$ for all $N\ge0$, without training and with $B=0$.
\item[(ii)] If $w<1$ and $Mw<1$, then $R_N=1$ for all $N$: some world is served by no state.
\item[(iii)] If $w<1\le Mw$, then $M\ge2$. With the covering margin
\[
d_*=\frac{Mw-1}{M-1}\in[0,w),
\]
we have, for every $N\ge1$,
\[
\frac2M\,\Phi\!\Bigl(-\frac{\sqrt N\,d_*}{2\sigma}\Bigr)\ \le\ R_N\ \le\ \min\Bigl\{1,\ 2\Phi\!\Bigl(-\frac{\sqrt N\,d_*}{2\sigma}\Bigr)\Bigr\}.
\]
If $d_*>0$, the exact reliability exponent is $\lim_N-\frac1N\log R_N=d_*^2/(8\sigma^2)$, and the number of measurements needed for $R_N\le\delta$ is $(8\sigma^2/d_*^2)\log(1/\delta)\,(1+o(1))$ as $\delta\downarrow0$ at fixed $M$.
\end{enumerate}
The upper bound is attained by $M$ equally spaced windows of length $w$ that overlap by exactly $d_*$, with a midpoint-threshold encoder of $\lceil\log_2M\rceil$ bits. Each window is tiled by $\Leq$ candidate intervals, of which the first $\Leq-1$ are verified and the last is delivered.
\end{theorem}

\paragraph{Why capacity sets the learning rate.} A state can serve an interval of length $h$ of plausible parameters only if its success set contains it. If the success set has total length at most $w$, the left endpoints of such intervals form a set of measure at most $(w-h)_+$; all left endpoints in $[0,1-h]$ must be served, so $M(w-h)\ge1-h$, that is, $h\le d_*$. Under Gaussian measurement an interval of length $h$ has common difficulty $h^2/(8\sigma^2)$, so the exponent is the common difficulty of the largest commonly coverable interval. Theorem~\ref{thm:coverage} does not apply to the continuum; the proof (Appendix~\ref{app:continuous}) uses an interval-erosion lemma.

\begin{corollary}[Continuous coverage number and the exchange of bits for verification]\label{cor:contK}
Under the contract of Theorem~\ref{thm:continuous} with $w<1$, let $\gamma>0$ and $h_\gamma=\sigma\sqrt{8\gamma}$. The least number of addressable states achieving reliability exponent $\gamma$ is
\[
K^{\mathrm{cont}}_\gamma=\Bigl\lceil\frac{1-h_\gamma}{w-h_\gamma}\Bigr\rceil\quad\text{if }h_\gamma<w,\qquad K^{\mathrm{cont}}_\gamma=\infty\quad\text{if }h_\gamma\ge w .
\]
Hence $\lceil\log_2K^{\mathrm{cont}}_\gamma\rceil$ readable bits are necessary and sufficient. Each extra verified candidate widens $w$ by $2\tau$; when $h_\gamma\ll w$, doubling $\Leq$ saves about one bit, and no capacity buys an exponent above $w^2/(8\sigma^2)$.
\end{corollary}

Sharp cases ($Mw=1$, $\Leq=1$, $M=2$), a weak-signal comparison and a binary-measurement bridge are in Appendix~\ref{app:extgeo}.

\subsection{Discrete worlds: the covering order}\label{sec:discrete}

\paragraph{Contract.} The true rule is $\theta\in[K]$. Each document carries a token naming the true rule with probability $p$ and each other rule with probability $q=(1-p)/(K-1)<p$. The task is to deliver the answer of $\theta$ under an equality verifier. A public partition splits $[K]$ into $2^e$ cells of size $m=K/2^e$, and $e$ evidence bits reveal the cell of $\theta$ (an index oracle, departing from (C6)). Deployment is deterministic and probes at most $c$ of $B$ frozen bits, so within a cell it distinguishes at most $M=2^{\min(B,c)}$ states, each serving at most $\ell=\min(m,g,v+1)$ rules per cell. Let $C(m,\ell,t)$ be the least number of $\ell$-subsets of an $m$-set containing every $t$-subset \citep{gordon1995covering}, $C(m,\ell,0)=1$, and define the \emph{covering order}
\[
t_*=\max\{0\le t\le\ell:\ C(m,\ell,t)\le M\},\qquad s=t_*+1,
\]
and $E_s=-\log\bigl[(K-s)q+s\,p^{1/s}q^{1-1/s}\bigr]$.

\begin{theorem}[Discrete covering order]\label{thm:discrete}
Let $g\ge1$. If $\ell=m$, then $R_N=0$ for every $N\ge0$. Otherwise, for every $N\ge1$:
\begin{enumerate}
\item[(i)] every scheme satisfies $R_N\ge\frac1s\int\min_{i\in S}P_i^{\otimes N}$ for some set $S$ of $s$ rules in one cell. In particular $R_N=1$ if $t_*=0$, and $-\frac1N\log R_N\le E_s+O(\frac{\log N}{N})$;
\item[(ii)] if $t_*\ge1$, the following algorithm achieves $R_N\le\binom{K-1}{s-1}e^{-NE_s}$. Count token frequencies; store a covering block that contains the within-cell indices of the $t_*$ most frequent rules; at deployment, read the cell, generate the block's $\ell$ candidates in that cell, verify $\ell-1$ of them, and deliver the last.
\end{enumerate}
Hence $\lim_N-\frac1N\log R_N=E_s$. The exponent $E_s=C(S)$ is the common difficulty of any $s$ rules.
\end{theorem}

\paragraph{Relation to the coverage theorem.} Here $C(H)$ depends only on $|H|$, so $\cH_\gamma$ consists of all sets of size at most $t$ when $E_t<\gamma\le E_{t+1}$, and covering them with states serving at most $\ell$ rules per cell needs exactly $C(m,\ell,t)$ states. Thus $K_\gamma=C(m,\ell,t)$ and Theorem~\ref{thm:discrete} is Theorem~\ref{thm:coverage} with the covering number in closed form, plus an explicit algorithm. With $K=3$, $p=0.8$ and $\ell=2$, one readable bit gives $E_2=0.4069$, two bits give $E_3=0.5108$, and no bits give persistent failure: capacity changes the order of the hardest set to be separated, and hence the learning rate. In both geometries the exponent is the common difficulty of the smallest set that cannot be covered; the proofs differ, and neither formula is a substitution instance of the other.

\section{Resource conversion}\label{sec:conversion}

Learning supplies a program; deployment must still handle facts that were never seen. Suppose the regularity $F$ is known. How far can generation and verification replace reading fresh facts?

\paragraph{Contract.} A task presents fresh facts $x\in[0,1]^p$, oriented and scaled coordinatewise. The known program $F:[0,1]^p\to\R$ is continuous and coordinatewise nondecreasing, and has \emph{increasing differences}: for every split of coordinates into $J$ and its complement, the increment $F(t,z)-F(t,y)$ with $y\le z$ on the complement is nondecreasing in $t$ on $J$. A twice differentiable $F$ has this property if all mixed second derivatives are nonnegative. Reading a coordinate returns its exact value. The set of coordinates read on any path must belong to a public family $\cA$ that contains $\emptyset$ and is closed under subsets, for example $\{J:|J|\le e\}$. The controller may generate $g$ candidates and query an exact radius-$\tau$ verifier $v$ times. It must deliver a generated $u$ with $|u-F(x)|\le\tau$ for \emph{every} $x$; this is a deterministic worldwise guarantee in exact real arithmetic, with unlimited planning computation. Define
\[
D_{\cA}(F)=F(\mathbf1)-\max_{J\in\cA}F(\mathbf1_J),\qquad \nu=\min(g,v+1),
\]
where $\mathbf1_J$ is the indicator vector of $J$.

\begin{theorem}[Resource conversion]\label{thm:conversion}
Under this contract, a deterministic controller with a worldwise guarantee exists if and only if
\[
g\ge1\qquad\text{and}\qquad D_{\cA}(F)\le2\tau\,\min(g,v+1).
\]
When the condition holds, a non-adaptive controller suffices: read a maximizer $J^*$ of $F(\mathbf1_J)$, split the remaining output interval into $\nu$ pieces, verify the centers of the first $\nu-1$, and deliver the last.
\end{theorem}

\paragraph{The common worst corner.} After reading $x_J=t$, the outputs fill $[F(t,\mathbf0),F(t,\mathbf1)]$, longest at $t=\mathbf1$ for every $J$. Answering every read with $1$ and every verification with ``no'' leaves outputs filling an interval of length $D_{\cA}(F)$, while the controller succeeds on at most $\nu$ intervals of radius $\tau$. Each verified candidate buys $2\tau$ of tolerated uncertainty.

\begin{corollary}[Conversion margins define faces of the success complex]\label{cor:faces}
Let $\{F_\theta\}_{\theta\in\Theta}$ be a finite family of continuous programs, one for each world. The fresh facts $x$ are task randomness, and the target is worldwise success for every $x$. Each deployment policy reads coordinates in $\cA$, generates candidates and verifies them, as in Theorem~\ref{thm:conversion}.
\begin{enumerate}
\item[(i)] Suppose there is a center $\widehat F$ satisfying the hypotheses of Theorem~\ref{thm:conversion} such that $\sup_{\theta\in H}\|F_\theta-\widehat F\|_\infty\le\eta$ and $D_{\cA}(\widehat F)+2\eta\le2\tau\nu$. Then $H$ is compatible: one policy serves every world of $H$ for every $x$.
\item[(ii)] If some $\theta\in H$ has $F_\theta$ satisfying the hypotheses and $D_{\cA}(F_\theta)>2\tau\nu$, then $\{\theta\}$, and hence $H$, is incompatible.
\item[(iii)] Consequently, if $\{P_\theta\}$ satisfies the hypotheses of Theorem~\ref{thm:coverage} and every member of $\cH_\gamma$ lies in one of $M'$ sup-norm balls whose centers and radii satisfy (i), then $K_\gamma\le M'$, and Theorem~\ref{thm:coverage}(a) gives exponent $\gamma$ with $\lceil\log_2M'\rceil$ frozen bits.
\end{enumerate}
\end{corollary}

The conversion margin $2\tau\nu-D_{\cA}(\widehat F)$ is the radius a ball of learned programs may have, so generation and verification enlarge the faces of the success complex. Appendix~\ref{app:extconv} gives the value of adaptivity without increasing differences (Proposition~\ref{prop:certificates}), the knowledge-error band (Corollary~\ref{cor:band}), NP-completeness of choosing evidence (Proposition~\ref{thm:planning}) and randomization.

\section{A sharp end-to-end theorem}\label{sec:endtoend}

We now join learning and deployment in one problem with noisy experience about many shared primitives, an unseen nonlinear combination, frozen knowledge, independent external facts, local reads, generation and ordered comparison. Its target is oracle-relative, and its success complex is computed explicitly.

\subsection{Model and success geometry}

\paragraph{Contract.} Fix $d\ge k\ge2$, $\rho\in(0,1]$ and $\eps\in(0,\frac12)$, all public; put $\lambda=\log\frac{1-\eps}{\eps}$ and $a_\eps=2\sqrt{\eps(1-\eps)}\in(0,1)$. The world is $\theta\in\{0,1\}^d$.
\begin{itemize}
\item \emph{Documents.} Each carries a uniform address $i\in[d]$ and, with probability $\rho$, the token $\theta_i$ flipped with probability $\eps$; otherwise it is erased. Write $n=\min(N,T_{\mathrm{read}})$.
\item \emph{Tasks.} A task is an adversarial set $I\subset[d]$ of $k$ addresses and a uniform public polarity $m\in\{0,1\}^I$. Put $J=\sum_{i\in I}(\theta_i\oplus m_i)$ and $f(j)=\binom j2$. Training never shows two addresses together.
\item \emph{Fresh facts.} Independent uniform integers $Z_E\in[0,2^h)$ (external) and $Z_C\in[0,2^b)$ (local); the target is $Y=f(J)+2^bZ_E+Z_C$. Evidence and local reads reveal bits of $Z_E$ and $Z_C$ from the most significant downward. With $e'=\min(e,h)$, $c'=\min(c,b)$, the residual offset after reading both prefixes is uniform on
\begin{equation}\label{eq:residual}
\cD=\{2^b\alpha+\beta:\ 0\le\alpha<2^{h-e'},\ 0\le\beta<2^{b-c'}\},\qquad u=|\cD|=2^{h+b-e'-c'}.
\end{equation}
\item \emph{Generation and ordered verification.} Per path, at most $g$ integers are generated and at most $v$ of them are compared with $Y$ ($<,=,>$); success means delivering $Y$.
\end{itemize}
Let $L=2^{\min(g,v+1)}-1$ be the comparison capacity of Lemma~\ref{lem:capacity}(c) ($L=0$ if $g=0$) and $A_j=f(j)+\cD$, $0\le j\le k$.

\begin{corollary}[Compatibility of translated supports]\label{cor:compat}
A set $\Gamma$ of activation counts admits one controller achieving the oracle value $\min(1,L/u)$ under every $j\in\Gamma$ if and only if $|\bigcap_{j\in\Gamma}A_j|\ge L$ when $L\le u$, and $|\bigcup_{j\in\Gamma}A_j|\le L$ when $L\ge u$.
\end{corollary}

\begin{definition}[Critical resolution scale]\label{def:sstar}
A \emph{window} is a set $\{t,t+1,\dots,t+r\}\subseteq\{0,\dots,k\}$ of span $r$. Let $s_*$ be the least $r\in\{1,\dots,k\}$ for which some window of span $r$ violates the criterion of Corollary~\ref{cor:compat}, and $s_*=\infty$ if none does.
\end{definition}

The scale $s_*$ is computable from $(k,\cD,L)$ alone. Flipping one primitive changes $J$ by at most one, so world sets of Hamming diameter below $s_*$ are compatible, while a flip path across an incompatible window is not; incompatibility can have order $2^{\nu-1}$ (Proposition~\ref{prop:highorder}).

\subsection{The exact frontier}\label{sec:frontier}

Under the uniform prior the posterior of $\theta$ given the corpus is a product of Bernoulli laws: with $\mathsf s_i$ the number of tokens $1$ minus the number of tokens $0$ at address $i$, $\Prob(\theta_i=1\mid\text{corpus})=\sigma(\lambda\mathsf s_i)$, where $\sigma(z)=1/(1+e^{-z})$. For a task $(I,m)$ put $p_i=\sigma(\lambda\mathsf s_i)$ if $m_i=0$ and $p_i=1-\sigma(\lambda\mathsf s_i)$ if $m_i=1$, let $(q_j)_{j=0}^k$ be the coefficients of $\prod_{i\in I}(1-p_i+p_iz)$, and define the posterior law of the residual target
\begin{equation}\label{eq:mu}
\mu_n(y)=\frac1u\sum_{j=0}^k q_j\,\Ind\{y-f(j)\in\cD\},\qquad y\in\Z .
\end{equation}
The \emph{posterior controller} reads the $e'$ external and $c'$ local prefix bits, subtracts the known shift, sorts the $L$ heaviest values of $\mu_n$ (ties broken by integer value), and runs the balanced comparison tree of Lemma~\ref{lem:capacity}(c) on them.

\begin{theorem}[Exact frontier]\label{thm:frontier}
For all $n\ge0$ and budgets $(e,c,g,v)$:
\begin{enumerate}
\item[(i)] $\Vor=\min(1,L/u)$ for every $\theta$ and every task.
\item[(ii)] For $B\ge\min\{d\lceil\log_2(2n+1)\rceil,\ n\lceil\log_2(2d+1)\rceil\}$ (enough to store the counts or the corpus itself) and sufficient workspace and arithmetic,
\[
V^*_n(e,c,g,v)=\E\,\Top_L(\mu_n),
\]
the expectation being over the corpus and polarity under any fixed $\theta$ (equivalently, under the uniform prior). The posterior controller attains $V^*_n$ with success probability that is the same for every $\theta$ and every $I$.
\item[(iii)] The upper bound $V^*_n\le\E\,\Top_L(\mu_n)$ holds for every training rule reading at most $n$ documents, every frozen state, and every controller that interleaves evidence reads, local probes, generations and comparisons adaptively.
\end{enumerate}
\end{theorem}

The proof (Appendix~\ref{app:frontier}) reduces interleaved controllers to prefix readers, caps conditional success by $\Top_L$ (Lemma~\ref{lem:capacity}), and uses a label-flip symmetry from the public polarity to equate Bayes and worst-case values. Frozen-bit probes are unbudgeted here (Appendix~\ref{app:extend}).

\subsection{Sharp learning complexity}\label{sec:sharp}

Let $R_n=\Vor-V_n^*$. For $1\le s\le d$, let $\Omega_s(n)=\int\min(P_\theta^{n},P_{\theta'}^{n})$ for any $\theta,\theta'$ at Hamming distance $s$; by symmetry it depends only on $s$ and equals twice the optimal equal-prior error of testing $\theta$ against $\theta'$. With $\chi_s=\rho s/d$,
\begin{equation}\label{eq:omega}
\Omega_s(n)=2\,\E\Bigl[\Prob\bigl(\mathcal J<\tfrac{\mathcal M}{2}\bigm|\mathcal M\bigr)+\tfrac12\Prob\bigl(\mathcal J=\tfrac{\mathcal M}{2}\bigm|\mathcal M\bigr)\Bigr],
\end{equation}
where $\mathcal M\sim\Bin(n,\chi_s)$ counts documents that hit a differing address and $\mathcal J\mid\mathcal M\sim\Bin(\mathcal M,1-\eps)$ counts those supporting the first world.

\begin{theorem}[Finite-sample sandwich and exact exponent]\label{thm:sandwich}
If $s_*=\infty$, then $R_n=0$ for all $n\ge0$: one zero-data program attains the oracle. If $s_*<\infty$, then for every $n\ge0$,
\begin{equation}\label{eq:sandwich}
\frac{2^{-k}}{u\,s_*(s_*+1)}\,\Omega_{s_*}(n)\ \le\ R_n\ \le\ \binom{2^k}{2}\,\Omega_{s_*}(n),
\end{equation}
and consequently
\[
\lim_{n\to\infty}-\frac1n\log R_n=E_*:=-\log\Bigl[1-\frac{\rho s_*}{d}\bigl(1-2\sqrt{\eps(1-\eps)}\bigr)\Bigr].
\]
Both constants are independent of $d,\rho,\eps,n$. They depend on $k$ and $u$, and their ratio is of order $2^{3k}u\,s_*^2$, exponential in $k$ and in $h+b$. The lower bound holds for every training rule that reads at most $n$ documents, including rules that retain the whole corpus.
\end{theorem}

The exponent is the Bhattacharyya exponent of two worlds at distance $s_*$, in agreement with Theorem~\ref{thm:coverage}(d), since a critical flip chain has common difficulty exactly $E_*$ (Lemma~\ref{lem:chaindifficulty}). The lower bound uses a flip chain $\theta^0,\dots,\theta^{s_*}$ across a critical window and the following lemma.

\begin{lemma}[Cyclic rotation]\label{lem:cyclic}
For the chain above,
$\displaystyle \frac{1}{s_*}\,\Omega_{s_*}(n)\le\int\min_{0\le j\le s_*}P^n_{\theta^j}\le\Omega_{s_*}(n)$.
\end{lemma}

For each corpus some cyclic rotation of the flip order makes the chain minimum equal the endpoint minimum; averaging over rotations gives $1/s_*$ at every $n$ (Appendix~\ref{app:sandwich}).

\begin{theorem}[Second-order learning complexity]\label{thm:second}
Fix $k,h,b$ and $\eps\in(0,\frac12)$ and assume $s_*<\infty$. There are constants $0<c_1\le C_1$ depending only on $(k,h,b,\eps)$ such that for all $d\ge k$, $\rho\in(0,1]$, budgets with $s_*<\infty$, and $n\ge0$,
\[
c_1\,\frac{e^{-nE_*}}{\sqrt{1+nE_*}}\ \le\ R_n\ \le\ C_1\,\frac{e^{-nE_*}}{\sqrt{1+nE_*}} .
\]
Consequently the least $n$ with $R_n\le\delta$ satisfies, as $\delta\downarrow0$,
\[
n_\delta=\frac{\log(1/\delta)-\frac12\log\log(1/\delta)+O_{k,h,b,\eps}(1)}{E_*}+O(1),
\]
uniformly in $d$ and $\rho$; this is necessary and sufficient both for the number of available documents and for the read budget.
\end{theorem}

The proof (Appendix~\ref{app:second}) tilts each document to $\sqrt{P_\theta P_{\theta'}}/e^{-E_*}$ and applies lattice local bounds; the prefactor is classical at fixed parameters, its uniformity in $d,\rho$ is new.

\subsection{Compact frozen knowledge}\label{sec:compact}

Deployment only needs a program that is oracle-optimal on all plausible worlds, so it suffices to store a compatible set containing the truth with high probability.

\begin{theorem}[Compact frozen knowledge]\label{thm:compact}
Assume $s_*<\infty$ and put $r_*=s_*-1$, $V_{d,k}=\sum_{j=0}^{k-1}\binom dj$ and $\mu_\circ=\bigl(\binom dk2^{k+h+b}\bigr)^{-1}$.
\begin{enumerate}
\item[(i)] There is a public compiler from the counts $(\mathsf s_i)$ to a frozen code of at most
$B_{\mathrm{upper}}=d+(r_*+2)+2^{r_*+1}\bigl(\lceil\log_2(r_*+1)\rceil+r_*\lceil\log_2(d+1)\rceil\bigr)=d+O_{s_*}(\log(d+1))$ bits and a public deployment decoder whose regret $R^{\mathrm{cmp}}_n(\theta,I)$ satisfies, for every $n\ge0$,
\[
\frac{2^{-k}}{u\,s_*(s_*+1)}\Omega_{s_*}(n)\le\sup_{\theta,I}R^{\mathrm{cmp}}_n(\theta,I)\le \Vor\,\min\Bigl\{1,\ 2^{d-1}\sum_{i=s_*}^d\binom di\,\Omega_{s_*}(n)\Bigr\}.
\]
In particular the compact code has the optimal exponent $E_*$, and for fixed $d$ the rate and second-order expansion of Theorem~\ref{thm:second}.
\item[(ii)] For any training rule, any $B$-bit frozen code and any deployment rule (possibly randomized independently of $\theta$), if the worst-case regret is at most $\delta<\mu_\circ$ for some $n$, then $2^B\ge(1-\delta/\mu_\circ)\,2^d/V_{d,k}$. In particular, $R_n\to0$ requires $B\ge B_{\mathrm{lower}}=d-\log_2V_{d,k}$.
\end{enumerate}
\end{theorem}

The code stores a maximal set of diameter $s_*-1$, certified by at most $2^{s_*}$ Hamming balls \citep[Theorem~1.2]{alon2024helly}; necessity follows from Sauer--Shelah. Thus $d-\log_2V_{d,k}\le\log_2K_{0+}\le\log_2K_{E_*}\le B_{\mathrm{upper}}$ bounds the coverage number explicitly (Appendix~\ref{app:compact}).

\section{Unification and laws of solvability}\label{sec:laws}

Table~\ref{tab:quantities} lists the governing quantities. They are not values of one formula; each measures how statistically difficult world sets fit into the success regions that deployment resources provide, and each is computable before data are seen. Figure~\ref{fig:deps} (Appendix~\ref{app:extunif}) gives the dependency graph.

\begin{table}[t]
\centering\small
\begin{tabular}{@{}lp{0.48\linewidth}p{0.36\linewidth}@{}}
\toprule
quantity & what it governs & relation to Theorem~\ref{thm:coverage}\\
\midrule
$C(H)$ & difficulty of a world set; exponents are values of $C$ & definition (inherited)\\
$K_\gamma$ & capacity: $\gamma$ achievable iff $M\ge K_\gamma$ & the theorem\\
$d_*$ & exponent $d_*^2/(8\sigma^2)$; $K^{\mathrm{cont}}_\gamma$ (Cor.~\ref{cor:contK}) & same shape, separate proof\\
$t_*$ & exponent $E_{t_*+1}$; $K_\gamma=C(m,\ell,t)$ & instance\\
$D_{\cA}(F)$ & online solvability iff $D_{\cA}\le2\tau\nu$ (fresh facts, not worlds) & separate proof; faces of the complex (Cor.~\ref{cor:faces})\\
$s_*$ & $E_*=C(\text{critical chain})$; $R_n\asymp\Omega_{s_*}(n)$ & refinement\\
\bottomrule
\end{tabular}
\caption{Structural quantities of common-success coverage.}
\label{tab:quantities}
\end{table}

\paragraph{Answers to Q1--Q7.} Q1: exactly the distinctions between worlds that share a set of difficulty below $\gamma$ but no success region; in the models, $d_*$, $t_*$ or $s_*$. Q2: worlds inside a common success region, such as intervals shorter than $d_*$ or sets of diameter below $s_*$. Q3: frozen knowledge suffices when $D_{\cA}$ fits the candidate capacity, and knowledge error $\eta$ consumes $2\eta$ of it. Q4: when worlds share the complete observation record and need incompatible answers. Q5: through list capacities $\nu$ and $2^\nu-1$, which enter $w$, $\ell$, $2\tau\nu$ and $L$. Q6: budgets generate the menu and hence the success complex. Q7: $\lceil\log_2K_\gamma\rceil$ bits, with closed forms $K^{\mathrm{cont}}_\gamma$, $C(m,\ell,t)$ and $d\pm O(\log d)$.

\paragraph{Seven laws.} Each summarizes proved statements; Appendix~\ref{app:extunif} states proofs, assumptions and limiting counterexamples. (1)~\emph{Task-relevant knowledge:} only worlds that cannot share a target-attaining program must be separated, and incompatibility is higher-order (Theorem~\ref{thm:coverage}, Proposition~\ref{prop:highorder}). (2)~\emph{Deployment resources determine learning needs:} budgets reshape the success complex and hence $d_*$, $t_*$, $s_*$ (Proposition~\ref{prop:monotone}). (3)~\emph{Structure, not entropy, sets difficulty:} supports of equal size and entropy can have different $s_*$ and exponents (by example). (4)~\emph{Task-relevant compression:} $\lceil\log_2K_\gamma\rceil$ bits suffice (Theorems~\ref{thm:coverage}, \ref{thm:compact}). (5)~\emph{External information is irreplaceable} when records coincide and answers conflict (an elementary principle; Proposition~\ref{prop:barrier}). (6)~\emph{Feedback structure sets the value of search:} capacities $2^\nu-1$ versus $\nu$ (Lemma~\ref{lem:capacity}). (7)~\emph{Resource returns are joint:} one comparison raises absolute success at every $n$ yet roughly halves the regret exponent (Table~\ref{tab:example}).

\paragraph{Scope.} Theorem~\ref{thm:coverage} assumes finite world families, finite alphabets with common support and finite menus; infinite menus need a uniform failure margin. Section~\ref{sec:geometry} assumes the stated measurement models, exact verifiers and deterministic deployment. Theorem~\ref{thm:conversion} assumes continuity, monotonicity and increasing differences; without the latter adaptivity can help, and with discrete outputs the criterion is only sufficient. The end-to-end results assume uniform addresses, homogeneous noise, a public uniform polarity, a restricted combination family and ordered evidence streams. Address-dependent noise breaks the cyclic symmetry, and an unrestricted combination family admits worlds with identical corpora and opposite answers. Appendix~\ref{sec:scope} gives details.

\paragraph{Open problems.} Tractable structures for $K_\gamma$; infinite world classes; joint time--space--sample lower bounds; noisy or learner-dependent verifiers.

\clearpage
\bibliography{references}

\clearpage
\appendix
\section{Supplementary results and extended discussion}\label{app:extended}

This appendix collects the results and discussion that are referenced from the main text. Proofs are in Appendices~\ref{app:general}--\ref{app:endtoend}.

\subsection{Operational foundations}\label{app:extfound}

\paragraph{Five layers of indistinguishability.} The theory keeps five objects apart.
\begin{enumerate}
\item The \emph{true world} $\theta$.
\item \emph{Training-identifiable classes}: worlds with $P_\theta=P_{\theta'}$ cannot be told apart by any amount of experience. At finite $n$, the statistical difficulty of a set $H$ is measured by $C(H)$ (Section~\ref{sec:coverage}).
\item \emph{Frozen-distinguishable worlds}: those that induce different laws of $W$. Capacity $B$ and the access budget limit how many distinctions survive deployment.
\item \emph{Commonly servable sets}: sets of worlds on which one legal program attains the target.
\item \emph{Fresh facts}: independent of $\theta$, so known to no learner and to no oracle.
\end{enumerate}
Solvability is the interaction of these five layers.

\paragraph{Units.} Different sections instantiate the abstract budgets differently, and we never identify reads, probes of frozen bits and general computation. Table~\ref{tab:units} fixes the units.

\begin{table}[t]
\centering\small
\begin{tabular}{@{}p{0.17\linewidth}p{0.2\linewidth}p{0.2\linewidth}p{0.15\linewidth}p{0.14\linewidth}@{}}
\toprule
Section & experience & access budget $c$ & evidence $e$ & verifier\\
\midrule
\ref{sec:coverage} (general) & i.i.d.\ documents, finite alphabet & frozen-state addressing ($M$ states) & part of the program menu & any (via menu)\\
\ref{sec:continuous} (continuous) & $N$ Gaussian measurements & probes of frozen bits & none & radius $\tau$\\
\ref{sec:discrete} (discrete) & $N$ calibration tokens & probes of frozen bits & cell index bits & equality\\
\ref{sec:conversion} (conversion) & none ($F$ known) & --- & coordinate reads, family $\cA$ & radius $\tau$\\
\ref{sec:endtoend} (end-to-end) & $n$ document reads & sequential local input bits & ordered external bits & comparison $<,=,>$\\
\bottomrule
\end{tabular}
\caption{Units of the abstract resources in each section. Word operations (arithmetic, sorting, data structures) and workspace are always accounted separately from $c$, $B$ and $S$.}
\label{tab:units}
\end{table}

The proof (Appendix~\ref{app:capacity}) simulates the all-negative feedback path for (a) and (b). For (c) it defers each generation to its first use and solves the recursion $F(g,v)\le\max\{1,1+2F(g-1,v-1)\}$. A comparison splits the remaining candidates into two live subtrees, whereas an equality or radius answer ``no'' removes one candidate. This is why the capacities $2^\nu-1$ and $\nu$ differ even though each query returns at most $\log_23$ bits. Comparison feedback is used only in Section~\ref{sec:endtoend}. A verifier that also reported the sign of the error would allow bisection and change Sections~\ref{sec:continuous} and~\ref{sec:conversion}. Every theorem below states which interface it uses.

\subsection{Common-success coverage}\label{app:extcov}

\paragraph{Addressable states.} A scheme with $M$ \emph{addressable states} consists of a training rule mapping $X_{1:n}$ and internal randomness to $W\in[M]$, together with a public assignment $\phi_n:[M]\to\Pi_\omega$. A frozen state of $B$ bits addresses at most $2^B$ policies, so $M\le2^B$ always. A probe budget $c$ reduces this to $2^{\min(B,c)}$ only in special cases. If the contract has a single task and the verifier answers only ``yes'' or ``no'' (equality or radius feedback), the success set of a deterministic controller is determined by the path on which every answer is ``no''; along that path its probes form a decision tree with at most $2^c$ leaves, so at most $2^{\min(B,c)}$ distinct success sets occur (Appendix~\ref{app:continuous}). With several tasks, or with comparison feedback, the probed locations may depend on the task or on the verifier's answers. For example, with tasks $Q\in[B]$, worlds $\theta\in\{0,1\}^B$, target ``output $\theta_Q$'', $g=1$ and $c=1$, storing $W=\theta$ addresses all $2^B$ policies. In general we therefore treat $c$ as a restriction on the menu $\Pi_\omega$ and count $M\le2^B$. We allow $\phi_n$ to change with $n$, and allow deployment to randomize over policies. The \emph{failure probability} is $F_n(\theta)=\Prob_\theta(\theta\notin\Succ(\pi_W))$ and the risk is $R_n=\max_\theta F_n(\theta)$.

\paragraph{What the theorem says.} It answers Q1, Q2 and Q7 for finite problems. The learner must separate exactly those worlds that lie together in some set of difficulty below $\gamma$ but in no common success region. All other indistinguishability is harmless. Knowledge capacity enters only through the number of addressable programs, so the required number of bits, $\lceil\log_2K_\gamma\rceil$, can be far below $\log_2|\Theta|$, the cost of identification. Part (d) is the special case, for finite alphabets and in worst-case form, of the exponent of \citet[Theorem~2]{park2026regret}, who treat Bayesian problems with general observation spaces. Part (b) is a version of their common-mass argument, and the randomized constant $\eta_\Theta=(|\Theta|2^{|\Theta|})^{-1}$ comes from the same union control over subsets that gives their constant $|H|2^{|H|}$. The polynomial prefactor in (a) makes that bound loose at moderate $n$; it is the exponent and the capacity threshold that are sharp. The new content is the resource side. The menu $\Pi_\omega$ is generated by online budgets, frozen addressing caps the number of usable programs, and the theorem turns statistical difficulty into a knowledge-capacity requirement. A fixed menu has the exponent $\sup\{\gamma:\cH_\gamma\subseteq\bigcup_{w}2^{\Succ(\phi(w))}\}$, and optimizing the menu gives $E^*(M)$.

\paragraph{Relation to the finite decision-theoretic exponent.} With deterministic deployment and $M$ states, the learner faces an ordinary finite decision problem whose actions are the $M$ assigned policies. Applying \citet{park2026regret} to every $M$-subset of $\Pi_\omega$ and maximizing over subsets therefore gives the same value $E^*(M)$, in its Bayes form. We state this plainly: part (c) is not a new exponent principle. Its contribution is the covering formulation. That formulation reduces the capacity question to a single combinatorial number, $K_\gamma$, which the structured models of Sections~\ref{sec:geometry}--\ref{sec:endtoend} compute in closed form or bound sharply. It also handles randomized deployment. And it yields a witness bound valid at every $n$, not only an exponent.

The finiteness of the menu matters. With infinitely many programs, or with near-optimal policies whose success regions are not closed, part (b) needs a uniform failure margin. Section~\ref{sec:continuous} treats an infinite world family by a separate geometric argument.

\begin{proposition}[Oracle-relative targets]\label{prop:oracle}
Suppose each policy has values $V(\pi;\theta,Q)\in[0,1]$, the oracle value is $\Vor(\theta,Q)=\max_{\pi\in\Pi_\omega}V(\pi;\theta,Q)$, and $\Pi_\omega$ and $\cQ$ are finite. Define $\Succ(\pi)=\{\theta:V(\pi;\theta,Q)=\Vor(\theta,Q)\ \forall Q\}$ and let $\Delta>0$ be the least positive value of $\Vor-V$. Then the worst-case regret $R^{\mathrm{or}}_n$ satisfies $(\Delta/|\cQ|)\,R_n\le R^{\mathrm{or}}_n\le R_n$. Consequently the exponent and capacity statements of Theorem~\ref{thm:coverage}(a), (c) and (d) hold for oracle-relative regret unchanged, the constant $\eta_\Theta$ in (b) becomes $\eta_\Theta\Delta/|\cQ|$, and the bound in (e) becomes $R^{\mathrm{or}}_n\ge\Delta/(|\cQ|\,|H|)$.

Here the oracle is the best policy of the menu. It coincides with the same-budget oracle of Definition~\ref{def:frontier} when the menu contains every controller that respects $\omega$ and may depend on $\theta$; this holds in Section~\ref{sec:endtoend}, where the oracle-optimal lists are themselves menu elements.
\end{proposition}

\begin{proposition}[Computing the coverage number]\label{prop:setcover}
For an explicitly given finite menu, $K_\gamma$ is the set-cover number of the inclusion-maximal members of $\cH_\gamma$ by success regions. The greedy rule, which repeatedly adds the policy whose region contains the most uncovered maximal members, uses at most $(1+\ln|\cH_\gamma^{\max}|)\,K_\gamma$ policies. Deciding whether $K_\gamma\le M$ is NP-hard, already when $\cH_\gamma$ consists of singletons.
\end{proposition}

A greedy menu with $M'$ policies yields, through Theorem~\ref{thm:coverage}(a), a scheme with $\lceil\log_2M'\rceil$ bits, at most $\log_2(1+\ln|\cH^{\max}_\gamma|)+1$ bits above optimal. Set-cover hardness \citep{feige1998threshold} shows that, without structure, no polynomial-time procedure can approximate $K_\gamma$ much better. Structure is therefore what makes the capacity requirement predictable, and the models of Sections~\ref{sec:geometry}--\ref{sec:endtoend} supply it: there $K_\gamma$ is given by $\lceil(1-h_\gamma)/(w-h_\gamma)\rceil$, by $C(m,\ell,t)$, or by $d\pm O(\log d)$ bits.

Increasing online budgets enlarges the menu: $\omega\le\omega'$ componentwise implies $\Pi_\omega\subseteq\Pi_{\omega'}$. The effect on learning depends on the type of target.

\begin{proposition}[Resource monotonicity and its failure]\label{prop:monotone}
Let $\omega\le\omega'$.
\begin{enumerate}
\item[(i)] For a fixed target that does not depend on $\omega$, such as exact worldwise success, $\cS_\omega\subseteq\cS_{\omega'}$. Hence $K_\gamma(\omega')\le K_\gamma(\omega)$, $E^*(M;\omega')\ge E^*(M;\omega)$, and the capacity requirement $\lceil\log_2K_\gamma\rceil$ does not increase.
\item[(ii)] The absolute frontier $V^*$ is nondecreasing in every resource.
\item[(iii)] For oracle-relative targets the oracle value rises with $\omega$, success regions can shrink, and $E^*$ can decrease. In the end-to-end problem of Section~\ref{sec:endtoend}, one additional comparison raises absolute success at every sample size while roughly halving the regret exponent (Table~\ref{tab:example}).
\end{enumerate}
\end{proposition}

Part (iii) does not mean that verification makes a system less capable. The system becomes more capable; the reference it is measured against becomes more capable faster.

\begin{proposition}[Information barrier]\label{prop:barrier}
Fix a scheme, meaning a training rule together with a deployment policy, and two worlds $\theta,\theta'$. Suppose the complete observation record has the same law under $\theta$ and $\theta'$. The record consists of the corpus, all evidence and access returns, and all verifier answers. Suppose also that no delivered answer can succeed in both worlds for the same task realization. Then $V(\theta)+V(\theta')\le1$. This holds regardless of training size, frozen capacity, workspace and computation.
\end{proposition}

The delivered answer is a randomized function of the record, so it has the same law in both worlds; since the two success sets are disjoint, the two success probabilities sum to at most one. This settles Q4. Internal computation is a function of the record and cannot create the distinguishing fact. Verification is different: a verifier queried on a candidate that succeeds in one world but not the other answers differently, so it is itself a source of external information. With $v\ge1$ and $g\ge2$ the barrier can disappear. Theorem~\ref{thm:conversion} applies the barrier to fresh facts, and Theorem~\ref{thm:frontier}(i) to an oracle that knows $\theta$ but not the fresh facts. Theorem~\ref{thm:coverage}(e) is its training-time version.

\subsection{Geometries of solvability}\label{app:extgeo}

Corollary~\ref{cor:contK} is the continuous counterpart of $B_{\min}(\gamma)=\lceil\log_2K_\gamma\rceil$ in Theorem~\ref{thm:coverage}(c). Here the covering number has a closed form: the difficulty sets are intervals of length below $h_\gamma$, and $K^{\mathrm{cont}}_\gamma$ is the least number of success regions in which all of them fit.

\paragraph{Regimes.} The four cases behave differently. If $w\ge1$, no training is needed (Q2). If $Mw<1$, failure persists for any amount of data: capacity, not data, is the bottleneck (Q7). The critical case $Mw=1$ leaves a non-vanishing uniform risk, as the next proposition shows. If $Mw>1$, the overlap $d_*>0$ absorbs noise and the risk decays at rate $d_*^2/(8\sigma^2)$.

\begin{proposition}[Sharp cases]\label{prop:continuous-sharp}
Under the contract of Theorem~\ref{thm:continuous} with $w<1$:
\begin{enumerate}
\item[(i)] if $Mw=1$, then $R_N\ge\frac12$ for every $N$ and every $\Leq$;
\item[(ii)] if $\Leq=1$ and $Mw\ge1$, the lower bound improves to $R_N\ge\Phi(-\sqrt N d_*/(2\sigma))$;
\item[(iii)] if $M=2$ and $\frac12\le w<1$, then $R_N=\Phi(-\sqrt N(2w-1)/(2\sigma))$ exactly for every $N$ and every $\Leq$.
\end{enumerate}
\end{proposition}

\begin{proposition}[Same margin, different weak-signal risk]\label{prop:weak}
Let $M=2^b\ge4$ and $w=1-1/M$. Consider two contracts with the same $M$, $w$ and hence the same $d_*$ and exponent. In the first, $\Leq=M-1$ and $\tau=1/(2M)$, and the worst-case risk is at most $1/M$ for every $N$ and $\sigma$. In the second, $\Leq=1$ and $\tau=w/2$, and $\liminf_{\sigma\to\infty}R_N\ge\frac12$.
\end{proposition}

The comparison concerns the weak-signal envelope $\min\{1/M,\,2\Phi(\cdot)\}$, not exponents: the risk $1/M$ of the first contract does not vanish as $N$ grows, and both contracts have the same exponent.

Proposition~\ref{prop:weak} shows that the exponent is not a sufficient statistic for the whole finite-sample surface: how a success region is split into candidate intervals matters when the signal is weak.

\paragraph{Binary measurements.} The same deployment geometry governs a different statistical experiment. Replace Gaussian measurements by independent binary tokens with $\Prob(Y=1)=\frac14+\frac\theta2$.

\begin{corollary}[Measurement bridge]\label{cor:bridge}
Under the contract of Theorem~\ref{thm:continuous}(iii) with binary measurements, $\frac1{2M}e^{-4Nd_*^2/3}\le R_N\le\min\{1,2e^{-Nd_*^2/8}\}$ for all $N\ge1$. If $M=2$, $\Leq=1$ and $\frac12\le w<1$, then $R_N=\Prob(J>\frac N2)+\frac12\Prob(J=\frac N2)$ with $J\sim\Bin(N,\frac12-\frac{d_*}4)$ exactly, and $R_{2r}=R_{2r-1}$ for all $r\ge1$.
\end{corollary}

The two exponents in the corollary differ by the factor $32/3$, so the binary case is determined only to order $\Theta(d_*^2)$. The exact exponent is presumably a Chernoff quantity under optimized thresholds; midpoint thresholds are not optimal for Bernoulli measurements, and we do not compute it. The covering geometry is the same as in Theorem~\ref{thm:continuous}, while the finite-sample curves differ: the binomial risk has parity plateaus and the Gaussian risk does not. In both experiments the sample complexity is $\Theta(d_*^{-2}\log(1/\delta))$ at fixed $M$. The upper bound uses only the empirical frequency, which a single cross-entropy gradient step from zero also produces; no result depends on that implementation.

In both problems the governing quantity measures how large a statistically difficult set can be while still fitting inside one addressable success region. In the continuous problem difficulty is measured by length and coverage by overlap: $d_*$. In the discrete problem difficulty is measured by cardinality and coverage by covering designs: $t_*$. The exponent is in both cases the common difficulty of the smallest set that cannot be covered, $d_*^2/(8\sigma^2)$ and $E_{t_*+1}$ respectively. This is the shape predicted by Theorem~\ref{thm:coverage}. The proofs are different: one uses interval erosion and endpoint overlap, the other covering designs and multinomial types. Neither formula is obtained from the other by substitution.

\subsection{Resource conversion}\label{app:extconv}

\paragraph{What the theorem answers.} It gives exact exchange rates between resources (Q3--Q5).
\begin{itemize}
\item After the regularity is known, $D_{\cA}(F)$ measures exactly the residual output uncertainty caused by unread facts.
\item Each additional candidate together with one verification enlarges the tolerated uncertainty by $2\tau$.
\item Reading more facts lowers $D_{\cA}$ along the program's own geometry.
\item Uncertainty that execution can remove, the value of $F$ on read inputs, is separated from uncertainty that only observation can remove, the fiber of unread inputs.
\end{itemize}
The radius verifier here has capacity $\nu$, while the comparison verifier of Section~\ref{sec:endtoend} has capacity $2^\nu-1$. The feedback structure is part of the contract, not a detail.

\paragraph{Examples.} For a convex ridge $F(x)=\psi(\sum_ia_ix_i)$ with $a_i\ge0$ and $\psi$ convex and nondecreasing, increasing differences hold and the optimal $J$ reads the largest weights. For $F(x)=(\sum_ia_ix_i)^2$ with $a=(\frac78,\frac18,\frac18)$, one read gives $D=S^2-S_1^2=\frac{81}{64}-\frac{49}{64}=\frac12$, where $S$ is the sum of all weights and $S_1$ the largest weight. With $\tau=\frac18$, two candidates and one verification are exactly sufficient, and one candidate is not. Every polynomial with nonnegative coefficients satisfies the hypotheses on the box.

\begin{proposition}[Quantifier certificates and the value of adaptivity]\label{prop:certificates}
For any continuous $F$, let $o(J,z)$ be the oscillation of $F$ on the fiber $\{x:x_J=z_J\}$, and put
\[
\underline U=\sup_z\min_{J\in\cA}o(J,z),\qquad \overline U=\min_{J\in\cA}\sup_zo(J,z).
\]
If $g\ge1$, then $\underline U>2\tau\nu$ rules out a worldwise guarantee, and $\overline U\le2\tau\nu$ ensures one. Increasing differences give $\underline U=\overline U=D_{\cA}(F)$. Without increasing differences the gap can be strict and adaptivity can matter. For $F(t,x,y)=t+(1-t)x+ty$ on $[0,1]^3$, which is coordinatewise monotone but has $\partial_{tx}F=-1$, with two reads, $g=1$ and $v=0$, the best adaptive worst-case width is $\frac12$, every fixed pair of reads leaves width at least $1$, and $F(\mathbf1)-\max_{|J|\le2}F(\mathbf1_J)=0$ wrongly predicts width zero.
\end{proposition}

\begin{corollary}[Knowledge error converts into candidates]\label{cor:band}
Suppose the learner knows only an approximation $\widehat F$ with $\sup|\widehat F-F|\le\eta$, where $\widehat F$ satisfies the hypotheses of Theorem~\ref{thm:conversion} and $F$ is continuous. If $D_{\cA}(\widehat F)+2\eta\le2\tau\nu$ and $g\ge1$, a worldwise guarantee for $F$ exists. If $D_{\cA}(\widehat F)-2\eta>2\tau\nu$, none exists even for a controller that knows $F$ exactly.
\end{corollary}

Corollary~\ref{cor:band} connects learning to deployment in this setting. Knowledge precision $\eta$ costs $2\eta$ of tolerated width, that is, about $\eta/\tau$ additional verified candidates. Inside the band $-2\eta<D_{\cA}(\widehat F)-2\tau\nu\le2\eta$, deciding solvability requires more knowledge about $F$.

\begin{proposition}[Planning can be harder than execution]\label{thm:planning}
For a simple graph $G=([p],E)$ let $F_G(x)=\sum_{\{i,j\}\in E}x_ix_j$; it satisfies the hypotheses of Theorem~\ref{thm:conversion}. Evaluating $F_G$ at a rational input takes $O(p+|E|)$ arithmetic operations on numbers of polynomial bit length. Deciding, from the edge list, integers $e$ and $\nu$, and a rational $\tau$, whether a worldwise guarantee exists with $\cA=\{|J|\le e\}$ is NP-complete; this holds already for $g=1$, $v=0$.
\end{proposition}

Here $D_e(F_G)=|E|-\max_{|J|\le e}|E(G[J])|$, and the case $2\tau=|E|-\binom e2$ is the clique problem \citep{karp1972reducibility}; when $|E|=\binom e2$ this uses $\tau=0$, which the contract allows. Hardness is in the bit model, and ``harder than execution'' assumes $\mathrm{P}\ne\mathrm{NP}$. Choosing which facts to read can thus be computationally hard even when the regularity is known and executing it is linear. Planning costs, execution costs and compilation costs are separate resources. If a precomputed optimal evidence list is stored, its computation must be charged to compilation and its size to the frozen state.

\paragraph{Randomization.} Theorem~\ref{thm:conversion} concerns deterministic controllers with worldwise guarantees. If deployment may randomize after the evidence has been read, the worst-case success probability on a fixed slice of width $u-l$ equals $\min(1,\nu/K)$ with $K=\max(1,\lceil(u-l)/(2\tau)\rceil)$ (Appendix~\ref{app:conversion}). The deterministic threshold is the case $\nu\ge K$. Randomized choice of the evidence itself is not covered.

\subsection{The end-to-end model}\label{app:extend}

\begin{proposition}[High-order incompatibility]\label{prop:highorder}
Let $\cD=\{0,1\}$ (so $u=2$), $\nu=\min(g,v+1)\ge2$, $q=2^{\nu-1}$ and $k\ge q+1$. Then for every $(I,m)$, every set of at most $q-1$ worlds is compatible, some set of $q$ worlds is incompatible, and every incompatible set has at least $q$ worlds. Moreover $s_*=q-1$.
\end{proposition}

\begin{proof}
Each $A_j$ has two elements and $L=2q-1$. A set of at most $q-1$ worlds realizes at most $q-1$ counts, whose supports cover at most $2q-2<L$ integers; conversely an incompatible set needs a union exceeding $2q-1$, hence at least $q$ counts. For $j\ge2$, $f(j+1)-f(j)=j\ge2$, so $A_2,A_3,\dots,A_k$ are pairwise disjoint and the $q$ counts $k-q+1,\dots,k\ge2$ cover $2q>L$ integers. A window of span $q-1$ in $\{2,\dots,k\}$ is therefore incompatible, while every window of span $q-2$ contains $q-1$ counts.
\end{proof}

For $g=3$, $v=2$ ($L=7$) and $k=5$, the counts $2,3,4,5$ have answer sets $\{1,2\}$, $\{3,4\}$, $\{6,7\}$, $\{10,11\}$: any three fit into one seven-target search tree, all four do not. A pairwise analysis predicts zero regret; the true regret decays at the finite rate of Theorem~\ref{thm:sandwich} with $s_*=3$. The order of incompatibility grows exponentially with the verification depth.

\paragraph{Proof outline of Theorem~\ref{thm:frontier} (Appendix~\ref{app:frontier}).} Because both streams are ordered prefixes and budgets are per path, a controller that reads all $e'$ and $c'$ allowed bits first can simulate any interleaved controller. Given those prefixes the residual offset is uniform on $\cD$, independent of everything else. Lemma~\ref{lem:capacity} then caps conditional success by $\Top_L$ of the conditional target law; $\Top_L$ is convex, so coarsening the corpus to a frozen state cannot help, and the counts $(\mathsf s_i)_{i\in I}$ are sufficient. This yields the Bayes bound, which dominates the worst case. For equality, flipping all labels at address $i$ together with $m_i$ maps the joint law of (corpus, polarity) under $\theta$ to that under $\theta\oplus e_i$ while leaving $\mu_n$ and $Y$ unchanged; relabeling addresses does the same for $I$. Hence the posterior controller's success does not depend on $(\theta,I)$, and its Bayes value is its worst-case value. Without the uniform public polarity this argument gives only the Bayes value.

\paragraph{The sufficient statistic is one gradient step away.} Consider a predictor with one real parameter $w_i$ per address that assigns probability $\eps+(1-2\eps)\sigma(w_i)$ to token $1$ at an informative position of address $i$. At $w=0$ the gradient of the token cross-entropy is $(1-2\eps)(\frac12-y)$, and erased positions contribute nothing. One full-batch step on the summed loss with step size $2\lambda/(1-2\eps)$ yields $w_i=\lambda\mathsf s_i$, the posterior log-odds (Appendix~\ref{app:gradient}). Training sees only isolated addresses, while deployment reuses the same parameters for all $\binom dk$ query sets; a likelihood counter attains the same frontier.

\paragraph{Resources.} Training reads $n$ documents and performs $O(n+d)$ word operations; deployment uses $e'$ evidence reads, $c'$ local probes, at most $\min(g,v+1)$ generations and $\min(g,v+1)-1$ comparisons per path, and $O(k^2+(k+1)u\log((k+1)u+1))$ arithmetic operations on exact rationals (Proposition~\ref{prop:resources} in Appendix~\ref{app:frontier}). The budget $c$ counts input probes. Frozen-bit probes are unbudgeted in Section~\ref{sec:endtoend}: the posterior controller reads $k\lceil\log_2(2n+1)\rceil$ frozen bits, and the compact decoder of Section~\ref{sec:compact} reads $k$ bits of $\theta_0$ and its certificate. Charging these probes to $c$ would replace $c'$ by a smaller local budget and change the formula. The operation count includes the $O(k\log(n+1))$ word operations of repeated squaring, and counts word operations on integers of $\ell_\star$ bits, as in Proposition~\ref{prop:resources}. The factor $u=2^{h+b-e'-c'}$ is exponential in the length of the fresh input.

\paragraph{Discussion of Theorem~\ref{thm:sandwich}.} The exponent is the Bhattacharyya exponent $-\log(1-\chi_{s_*}(1-a_\eps))$ of two worlds at distance $s_*$: documents that do not hit a differing address have identical laws, and each hit is a binary symmetric observation. It is the exponent predicted by Theorem~\ref{thm:coverage}(d) with Proposition~\ref{prop:oracle} at fixed $d$: every incompatible set has diameter at least $s_*$ and therefore contains a pair with Chernoff information at least $E_*$, while a flip chain across a critical window has common statistical difficulty exactly $E_*$ (the normalized geometric mean of its endpoints is at equal divergence $E_*$ from every chain member; Lemma~\ref{lem:chaindifficulty}). What~\eqref{eq:sandwich} adds is control at every $n$ by a single binomial overlap, with constants that do not depend on $d$, $\rho$ or $\eps$.

\paragraph{Upper bound.} Fix $I,m$ and order the $2^k$ local worlds by likelihood. The longest prefix of diameter at most $s_*-1$ is compatible; if the true world lies outside it, its likelihood is at most the smaller likelihood of two prefix-breaking worlds at distance at least $s_*$. Summing over such pairs and using that $\Omega_{s}$ is nonincreasing in $s$ (thinning a more informative experiment by public erasures) gives the right side; the posterior controller is Bayes optimal and has constant regret, so it inherits the bound. Since regret never exceeds $\Vor=\min(1,L/u)$, the right side may be multiplied by $\Vor$.

\paragraph{Lower bound.} Fix a critical window $\{t,\dots,t+s_*\}$, the polarity $m=0$ (probability $2^{-k}$) and $s_*$ addresses $a_1,\dots,a_{s_*}$ of $I$. Starting from a world with $t$ ones in $I$, flipping $a_1,a_2,\dots$ in order gives a chain $\theta^0,\dots,\theta^{s_*}$ whose counts sweep the window. For every corpus, prefix and internal randomness, any controller's success set has at most $L$ elements, so by Corollary~\ref{cor:compat} the integer-valued deficits relative to the oracle sum to at least one over the chain; hence the chain-averaged regret is at least $2^{-k}\int\min_jP^n_{\theta^j}/(u(s_*+1))$. The remaining step compares the chain with its endpoints.

Lemma~\ref{lem:cyclic} is where multi-world incompatibility meets two-point testing: the chain is needed because pairs inside it may be compatible (Proposition~\ref{prop:highorder}), and keeping all $s_*+1$ worlds simultaneously likely costs only a factor $1/s_*$ at every $n$, including $n=0$ and majority ties. It uses exchangeability of the flipped addresses.

\paragraph{Discussion of Theorem~\ref{thm:second}.} The $O_{k,h,b,\eps}(1)$ term is of order $(3k+h+b)\log2$, so the $-\frac12\log\log(1/\delta)$ correction dominates it only when $\log\log(1/\delta)\gg k+h+b$. The prefactor $e^{-nE}/\sqrt n$ is classical at fixed parameters for lattice two-point testing \citep{bahadur1960deviations}; what is new is the uniformity in $d$ and $\rho$.

The proof (Appendix~\ref{app:second}) tilts each document to $\sqrt{P_\theta P_{\theta'}}/\varphi$, $\varphi=e^{-E_*}$. Under the tilt, the number of discriminative documents is $\Bin(n,z)$ with $z=\chi_{s_*}a_\eps/\varphi$ and the log-likelihood ratio is $\frac\lambda2$ times a simple random walk $\Sigma_K$, so $\Omega_{s_*}(n)=\varphi^n\,\E\,e^{-\lambda|\Sigma_K|/2}$ exactly. Lattice local bounds give $\E[e^{-\lambda|\Sigma_\ell|/2}]\asymp(1+\ell)^{-1/2}$ for all $\ell\ge0$ and both parities; averaging over $K$ and using $a_\eps\chi\le z\le\chi\asymp_\eps E_*$ yields the theorem. The constants degrade as $\eps\to\frac12$ (lattice spacing $\lambda\to0$) and as $\eps\to0$, where $\Omega_s(n)=(1-\chi_s)^n$ has no square-root prefactor. Since $E_*\asymp\rho s_*/d$, uniformity in $d$ says that $nE_*$, the effective number of discriminative documents, is the right clock for growing dimension. The compact code of Section~\ref{sec:compact} has a dimension-dependent constant.

\paragraph{Discussion of Theorem~\ref{thm:compact}.} For fixed $k$ both bounds are $d\pm O(\log d)$: about one bit per primitive, so for fixed $d$ the optimal exponent costs at most $O(\log d)$ more bits than vanishing regret. (The upper constant is of order $4^d$, so the compact guarantee becomes informative once $nE_*\gtrsim d$.) In the notation of Definition~\ref{def:coverage}, $d-\log_2V_{d,k}\le\log_2K_{0+}\le\log_2K_{E_*}\le B_{\mathrm{upper}}$. The exact finite-$n$ value $\E\,\Top_L(\mu_n)$ is guaranteed with the full counts; the compact code preserves the exponent.

\paragraph{Construction.} The majority vector $\theta_0$ maximizes the likelihood, and a world differing from $\theta_0$ on $F$ loses $\lambda\sum_{i\in F}|\mathsf s_i|$ in log-likelihood. Sort worlds by this loss (ties by $|F|$, then lexicographically) and take the longest prefix $H$ of Hamming diameter at most $r_*$; only the ball $B_{r_*}(\theta_0)$ and one outside witness (flip the $r_*+1$ smallest $|\mathsf s_i|$) need to be examined. Extend $H$ greedily inside $B_{r_*}(\theta_0)$ to an inclusion-maximal set $\mathcal C$ of diameter $r_*$. Maximality gives $\mathcal C=\bigcap_{y\in\mathcal C}B_{r_*}(y)$. By the Helly-type theorem of \citet[Theorem~1.2]{alon2024helly}, for $d>r_*$ some at most $2^{r_*+1}$ of these balls have the same intersection. The code stores $\theta_0$ and these centers as difference sets of size at most $r_*$. For a task $(I,m)$, the decoder enumerates the subsets of size at most $r_*$ of $I\cup R$, where $R$ is the union of the stored difference sets, and obtains the set of activation counts realized by $\mathcal C$ (removing coordinates outside $I\cup R$ never violates a stored constraint and does not change the count). This set lies in a window of span at most $r_*<s_*$, so Corollary~\ref{cor:compat} yields one oracle-optimal list, executed by Lemma~\ref{lem:capacity}(c). The decoder performs $O_{s_*}((k+1)^{r_*+1})$ word operations before list construction and reads only the $k$ bits of $\theta_0$ on $I$ and the certificate, not all $d$ stored bits. If $\theta\in H$ the regret is zero; otherwise $P_\theta^n(x)$ is at most the smaller likelihood of two worlds at distance at least $s_*$, which gives (i). A sunflower-based certificate gives the same order with explicit constants $j!(r_*+1)^j$ per layer (Appendix~\ref{app:compact}).

\paragraph{Necessity.} For a fixed deterministic deployment rule and code $w$, let $G_w$ be the worlds served at oracle level on every task. If $G_w$ shattered some $k$-set $I$, then for that $I$ every activation count in $\{0,\dots,k\}$ would be realized under a single program, contradicting the existence of an incompatible window; hence $\mathrm{VC}(G_w)\le k-1$ and $|G_w|\le V_{d,k}$ by the Sauer--Shelah lemma. The $2^B$ codes serve at most $2^BV_{d,k}$ worlds; every other world loses at least $\mu_\circ$ under uniform $(I,m)$, because for each task and each value of the read prefixes the success probability is a multiple of $1/u$. Averaging over a uniform world and over the deployment randomness gives (ii).

\subsection{Unification, laws, illustrations and scope}\label{app:extunif}

\paragraph{Computability.} All six quantities of Table~\ref{tab:quantities} can be computed before any data are seen, from the task, the deployment budgets and the measurement model: $K_\gamma$ by a covering computation over a finite menu, $d_*$ in closed form, $t_*$ from covering-design tables, $D_{\cA}(F)$ from two evaluations of $F$ per evidence set, and $s_*$ by a finite check of windows. The general computation of $K_\gamma$ is a set-cover problem and can take exponential time. In each structured model it has a closed form or a classical combinatorial value.

\begin{figure}[t]
\centering
\begin{tikzpicture}[
  box/.style={draw,rounded corners=2pt,align=center,font=\scriptsize,inner sep=3pt,minimum height=7mm,fill=white},
  >={Stealth[length=2mm]}, lab/.style={font=\tiny,fill=white,inner sep=1pt}]
\node[box] (mon) at (-3.6,4.6) {Prop.~\ref{prop:monotone}\\ resource monotonicity};
\node[box] (cov) at (0,4.6) {Theorem~\ref{thm:coverage}\\ coverage $\Leftrightarrow$ exponent};
\node[box] (orc) at (3.6,4.6) {Prop.~\ref{prop:oracle}\\ oracle-relative};
\node[box] (cap) at (-5.9,2.9) {Lemma~\ref{lem:capacity}\\ list capacity};
\node[box] (bar) at (5.9,2.9) {Prop.~\ref{prop:barrier}\\ information barrier};
\node[box] (u1) at (-5.0,1.2) {Thm.~\ref{thm:continuous}, Cor.~\ref{cor:contK}\\ margin $d_*$, $K^{\mathrm{cont}}_\gamma$};
\node[box] (u2) at (-1.9,1.2) {Thm.~\ref{thm:discrete}\\ order $t_*$};
\node[box] (c2) at (1.3,1.2) {Thm.~\ref{thm:conversion}, Cor.~\ref{cor:faces}\\ $D_{\cA}(F)\le2\tau\nu$};
\node[box] (fr) at (5.0,1.2) {Thm.~\ref{thm:frontier}\\ exact frontier};
\node[box] (br) at (-5.0,-0.5) {Cor.~\ref{cor:bridge}\\ binary measurement};
\node[box] (bd) at (-0.6,-0.5) {Cor.~\ref{cor:band}\\ knowledge band};
\node[box] (pl) at (2.4,-0.5) {Prop.~\ref{thm:planning}\\ planning hardness};
\node[box] (cy) at (6.1,-0.5) {Lemma~\ref{lem:cyclic}\\ cyclic rotation};
\node[box] (cp) at (-2.6,-2.2) {Thm.~\ref{thm:compact}\\ compact codes};
\node[box] (so) at (0.9,-2.2) {Thm.~\ref{thm:second}\\ second order};
\node[box] (sw) at (4.4,-2.2) {Thm.~\ref{thm:sandwich}\\ sandwich, $s_*$};
\draw[->] (cov) -- (mon); \draw[->] (cov) -- (orc);
\draw[->] (cov) -- node[lab]{instance} (u2);
\draw[->] (cap) -- (u1); \draw[->] (cap.east) to[out=0,in=135] (u2.north west);
\draw[->] (cap.east) to[out=-5,in=150] (c2.north west);
\draw[->] (bar) -- (c2); \draw[->] (bar) -- (fr);
\draw[->] (3.85,4.25) -- (3.85,-1.85);
\draw[->] (u1) -- (br); \draw[->] (c2) -- (bd); \draw[->] (c2) -- (pl);
\draw[->] (fr) -- (sw); \draw[->] (cy) -- (sw); \draw[->] (sw) -- (so); \draw[->] (sw) -- (cp);
\draw[->] (cap.south) to[out=-30,in=200] (fr.south west);
\draw[->,dashed] (cov.south west) to[out=200,in=60] (u1.north);
\draw[->,dashed] (cov.south) to[out=-80,in=110] (c2.north);
\draw[->,dotted,thick] (c2.north east) to[out=70,in=-40] (cov.south east);
\draw[->,dotted,thick] (cp.west) -| (-7.0,5.5) -| (cov.north);
\end{tikzpicture}
\caption{Dependency graph. Solid arrows are logical dependencies; the arrow labelled ``instance'' means the discrete theorem follows from the coverage theorem once $K_\gamma=C(m,\ell,t)$ is computed. Dashed arrows indicate that a result has the form predicted by the coverage theorem but needs its own proof (continuum; zero-data fresh-fact worlds). Dotted arrows indicate results that supply explicit bounds on the coverage number $K_\gamma$ (conversion margins as faces; compact codes).}
\label{fig:deps}
\end{figure}
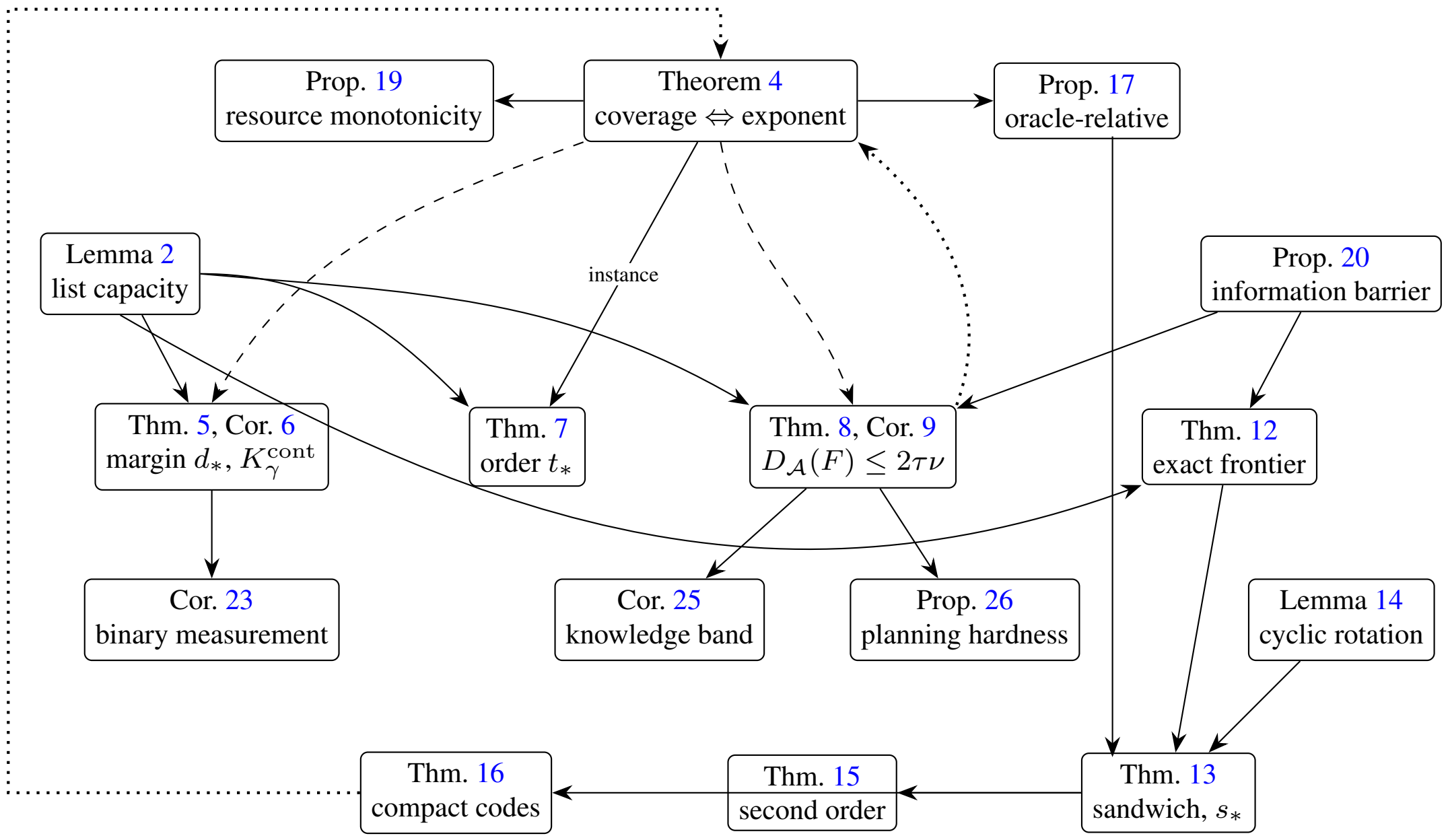

\paragraph{Answers to the seven questions.}

\begin{itemize}
\item \textbf{Q1 (what must be learned).} Exactly the distinctions between worlds that lie together in a set of difficulty below $\gamma$ but in no common success region (Theorem~\ref{thm:coverage}). In the structured models this means the order $t_*$, the margin $d_*$, or the resolution scale $s_*$.
\item \textbf{Q2 (harmless indistinguishability).} Worlds inside a common success region need not be separated. Examples: all of $\Theta$ when it is compatible (zero-data solvability); intervals shorter than $d_*$; sets of diameter below $s_*$, which the compact code stores without resolving.
\item \textbf{Q3 (when knowledge replaces observation).} Frozen knowledge suffices when the residual output uncertainty after evidence, $D_{\cA}$, fits the candidate capacity (Theorem~\ref{thm:conversion}). Knowledge error $\eta$ consumes $2\eta$ of this capacity (Corollary~\ref{cor:band}).
\item \textbf{Q4 (when computation is useless).} When the remaining worlds share the complete observation record and require incompatible answers (Proposition~\ref{prop:barrier}): fresh facts (Theorem~\ref{thm:frontier}(i)) and the common worst corner (Theorem~\ref{thm:conversion}).
\item \textbf{Q5 (generation and verification).} They act through list capacities: $\nu$ for equality and radius verifiers, $2^\nu-1$ for comparison (Lemma~\ref{lem:capacity}). These capacities enter $w=2\nu\tau$, $\ell$, $2\tau\nu$ and $L$.
\item \textbf{Q6 (deployment changes learning).} Online budgets generate the menu and hence the success complex (Proposition~\ref{prop:monotone}). This changes $d_*$, $t_*$ and $s_*$, and with them exponents and sample complexity.
\item \textbf{Q7 (minimal capacity).} $\lceil\log_2K_\gamma\rceil$ bits for exponent $\gamma$ (Theorem~\ref{thm:coverage}). Specific values: $\lceil\log_2K^{\mathrm{cont}}_\gamma\rceil$ with $K^{\mathrm{cont}}_\gamma=\lceil(1-h_\gamma)/(w-h_\gamma)\rceil$ (continuous, Corollary~\ref{cor:contK}); $C(m,\ell,t)$ (discrete); $d\pm O(\log d)$ (end-to-end, Theorem~\ref{thm:compact}).
\end{itemize}

\subsubsection{Seven laws}

Each law below summarizes proved statements. The assumptions and the counterexamples that limit it are stated with it.

\paragraph{Law 1 (task-relevant knowledge).} A system need not recover the world. It must separate only worlds that cannot share a target-attaining program. \emph{Proof:} Theorem~\ref{thm:coverage}(c,d) and Theorems~\ref{thm:continuous}, \ref{thm:discrete}, \ref{thm:sandwich}. \emph{Assumptions:} finite menus or the stated geometric contracts. \emph{Limit:} incompatibility is a higher-order relation. Proposition~\ref{prop:highorder} gives worlds of which every $2^{\nu-1}-1$ are compatible while some $2^{\nu-1}$ are not, so pairwise reasoning can predict zero error where the error is positive.

\paragraph{Law 2 (deployment resources determine learning needs).} Changing evidence, computation, generation or verification budgets changes the success complex and hence the knowledge that must be acquired. \emph{Proof:} Proposition~\ref{prop:monotone}; $d_*$ depends on $(M,\nu,\tau)$, $t_*$ on $(m,\ell,M)$, and $s_*$ on $(\cD,L)$. \emph{Example:} in Table~\ref{tab:example}, reading one external bit changes $s_*$ from $\infty$ to $2$ at $v=0$. With $e=1$, increasing $L$ through $1,3,7,15$ gives $s_*=2,1,2,\infty$, so verification can remove the need to learn altogether.

\paragraph{Law 3 (structure, not entropy, determines difficulty).} \emph{Status: established by examples within the models.} Uncertainties of equal size, equal entropy, or even equal exponent can have different learning costs. \emph{Proof:} in the end-to-end model with $h=1$, $b=2$, $k=4$, $L=1$, the residual supports $\{0,1,2,3\}$ and $\{0,1,4,5\}$ both have $u=4$ and two bits of entropy, but $s_*=2$ versus $s_*=1$ and the exponents differ by a factor of about two. Proposition~\ref{prop:weak} gives two contracts with the same $d_*$ and exponent whose weak-signal risks are $1/M$ and at least $\frac12$.

\paragraph{Law 4 (task-relevant compression).} Optimal frozen representations keep the distinctions the task needs, not the world. \emph{Proof:} $B_{\min}(\gamma)=\lceil\log_2K_\gamma\rceil$ (Theorem~\ref{thm:coverage}). The continuous problem needs $\lceil\log_2M\rceil$ bits for a parameter that would require unbounded bits to identify. The end-to-end compact code stores a maximal plausible set of diameter $s_*-1$ in $d+O(\log d)$ bits and is exponent-optimal (Theorem~\ref{thm:compact}). \emph{Limit:} our exactness result at finite $n$ uses enough capacity to store the counts or the corpus, and we prove no matching lower bound; the compact code preserves the exponent, not the finite-sample optimum.

\paragraph{Law 5 (external information is irreplaceable).} \emph{Status: an elementary principle (data processing), sharpened by the theorems cited.} If worlds that agree on all experience and all legal observations require incompatible answers, no internal computation solves the task in all of them. \emph{Proof:} Proposition~\ref{prop:barrier}; Theorem~\ref{thm:coverage}(e) (persistent failure); Theorem~\ref{thm:conversion} (worst-corner fiber); Theorem~\ref{thm:frontier}(i), which gives $V^*\le\min(1,L/u)$ for all training resources. \emph{Qualification:} true verification is itself an external channel. It can separate such worlds, and in the end-to-end model raising $L$ to $|\bigcup_jA_j|$ removes the need for training (Table~\ref{tab:example}, $s_*=\infty$). Evidence cannot replace training at a \emph{fixed} deployment budget with $s_*<\infty$, since then $R_0>0$; larger budgets can change this.

\paragraph{Law 6 (feedback structure sets the value of search).} \emph{Status: an elementary principle, made exact by Lemma~\ref{lem:capacity}.} The value of verification depends on how its answers partition the candidate space and on whether the program can use the partition, not only on the number of bits returned. \emph{Proof:} Lemma~\ref{lem:capacity}: comparison gives capacity $2^\nu-1$, equality and radius verification give $\nu$, with the same number of queries. A sign-revealing radius verifier would allow bisection and change Theorems~\ref{thm:continuous} and~\ref{thm:conversion}.

\begin{figure}[t]
\centering
\includegraphics[width=0.82\linewidth]{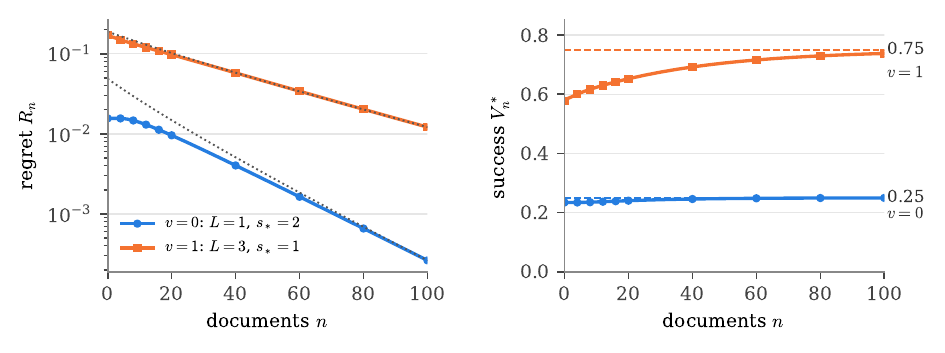}
\caption{End-to-end model with $e=1$, $g=3$: exact $R_n$ and $V^*_n$ for $n\le100$. Dotted lines show $Ce^{-nE_*}/\sqrt{1+nE_*}$, matched at $n=100$. One comparison ($v=1$) triples the oracle value and raises absolute success at every $n$, yet roughly halves the regret exponent. The sandwich~\eqref{eq:sandwich} holds at every computed $n$.}
\label{fig:frontier}
\end{figure}

\paragraph{Law 7 (resource returns are joint).} Adding one resource changes the marginal value of others. Absolute success and oracle-relative regret can move in opposite directions. \emph{Proof:} Proposition~\ref{prop:monotone}(ii,iii); Table~\ref{tab:example} and Figure~\ref{fig:frontier}: one comparison raises absolute success at every $n$ while roughly halving the exponent of convergence to the stronger oracle. In Theorem~\ref{thm:continuous}, frozen capacity changes the margin $d_*$ and hence the value of each additional measurement. \emph{Interpretation:} the decrease concerns convergence toward a stronger reference. It does not mean the system becomes less capable.

\subsubsection{Quantitative illustrations}

\begin{table}[t]
\centering\small
\begin{tabular}{@{}ccccccccc@{}}
\toprule
$e$ & $v$ & $u$ & $L$ & $\Vor$ & $V^*_6$ (exact) & $R_6$ & $s_*$ & $E_*$\\
\midrule
0 & 0 & 8 & 1 & 0.125 & 0.125000 & 0 & $\infty$ & ---\\
0 & 1 & 8 & 3 & 0.375 & 0.369037 & 0.005963 & 3 & 0.066469\\
0 & 2 & 8 & 7 & 0.875 & 0.795184 & 0.079816 & 1 & 0.021669\\
0 & 3 & 8 & 15 & 1 & 1 & 0 & $\infty$ & ---\\
1 & 0 & 4 & 1 & 0.25 & 0.234671 & 0.015329 & 2 & 0.043818\\
1 & 1 & 4 & 3 & 0.75 & 0.608910 & 0.141090 & 1 & 0.021669\\
1 & 2 & 4 & 7 & 1 & 0.957415 & 0.042585 & 2 & 0.043818\\
1 & 3 & 4 & 15 & 1 & 1 & 0 & $\infty$ & ---\\
\bottomrule
\end{tabular}
\caption{End-to-end model with $d=5$, $k=4$, $\rho=\frac45$, $\eps=\frac14$, $h=3$, $b=0$, $g=4$. Success at $n=6$ is computed exactly from $\E\,\Top_L(\mu_6)$ in rational arithmetic (Appendix~\ref{app:numerics}). $E_*$ is the limit in Theorem~\ref{thm:sandwich}, not a fit.}
\label{tab:example}
\end{table}

Four further instances are verified exactly or numerically in Appendix~\ref{app:numerics}. In a four-world menu in which every pair, but no triple, is compatible, the optimal exponent rises from $0$ with one state to $0.039$ with two and $0.071$ with three or more, as predicted by Theorem~\ref{thm:coverage}; the computed Bayes-risk slopes are $0.041$ and $0.072$. The discrete example of Section~\ref{sec:discrete} has exponents $0.407$ with one readable bit and $0.511$ with two. In the continuous problem, $2{,}137$ random menus never exceed the margin $d_*$, and the window construction attains it. Exhaustive decision programs confirm the worst-corner formula, the adaptivity gap of Proposition~\ref{prop:certificates}, and the clique reduction.

\subsubsection{Scope and open problems}\label{sec:scope}

\paragraph{What the theory assumes.} Theorem~\ref{thm:coverage} assumes finite world families, finite alphabets with common support, and finite program menus. A uniform positive failure margin is essential: with infinite menus or near-optimal policies the converse needs an additional argument, as in Section~\ref{sec:continuous}. The continuous and discrete theorems assume the stated measurement models, exact radius or equality verifiers, and deterministic deployment; training may be randomized. Theorem~\ref{thm:conversion} assumes continuity, monotonicity and increasing differences, exact reads, and exact real arithmetic with unlimited planning. Without increasing differences adaptivity can help (Proposition~\ref{prop:certificates}), and with discrete outputs the continuous criterion is only sufficient. The end-to-end results assume uniform addresses, homogeneous noise, a public uniform polarity, a public combination rule $f$, ordered evidence streams and an exact comparator. Without the polarity the Bayes value remains an upper bound but the worst case may differ; address-dependent noise breaks the cyclic symmetry; arbitrary-address reads break the prefix reduction. If the combination family is not restricted, an interaction term that vanishes on single addresses yields two worlds with identical corpora and opposite answers. Word-operation counts are sufficient budgets, not lower bounds. We do not determine a complete eight-resource exact frontier for any model. In the end-to-end model the frontier is exact in $(N,T_{\mathrm{read}},e,c,g,v)$ at sufficient frozen capacity and workspace, with frozen probes unbudgeted; below that capacity only the exponent is determined, and the dependence on workspace $S$ is not determined. Training workspace $S$ and training operations $T_{\mathrm{ops}}$ never bind in our results except through planning hardness (Proposition~\ref{thm:planning}).

\paragraph{Open problems.}
\begin{enumerate}
\item Computing $K_\gamma$ is a set-cover problem over the success complex. Which structural properties of a deployment interface make it tractable, and when is the reduction from complex to cover number lossy?
\item Extend Theorem~\ref{thm:coverage} to infinite world classes with uniform margins, unifying the interval-erosion argument of Section~\ref{sec:continuous} with metric-entropy methods.
\item Find finite-sample curves beyond the exponent in growing dimension for compact codes, and the dimension dependence of the constant in Theorem~\ref{thm:compact}(i).
\item Determine joint time--space--sample lower bounds for the whole pipeline. Our word-operation counts are only sufficient, and Proposition~\ref{thm:planning} shows that planning can dominate execution.
\item Extend the theory to noisy or self-referential verifiers, whose answers may depend on the learner; the all-negative path counting of Lemma~\ref{lem:capacity} must then be rebuilt.
\item Characterize how particular learning mechanisms, such as shared parameters or sequence prediction, realize or alter the coverage quantities defined here. This is the subject of separate work, and no result here depends on it.
\end{enumerate}
\section{Theorem-level attribution and related work}\label{app:attribution}

\begin{table}[h]
\centering\scriptsize
\begin{tabular}{@{}p{0.34\linewidth}p{0.32\linewidth}p{0.26\linewidth}@{}}
\toprule
Result & Inherited ingredient & Status here\\
\midrule
Lemma~\ref{lem:minimax}: $C(H)=\min_Q\max_\theta\KL=$ multivariate Chernoff information & classical minimax identity; multivariate Chernoff information \citep{mishra2024antidistinguishability} and the divergence-radius form in hypothesis exclusion \citep{ji2025exclusion}; pairwise $M$-ary exponents \citep{leang1997asymptotics} as context & proved for completeness\\
Lemma~\ref{lem:types}: common mass via types & method of types \citep{csiszar2011information} & classical\\
Theorem~\ref{thm:coverage}(d): exponent $=\min C$ over minimal incompatible sets & \citet[Thm.~2 and Sec.~V]{park2026regret} & inherited; special case (finite alphabet, worst-case risk) of their result\\
Theorem~\ref{thm:coverage}(a)--(c),(e): exponent $\gamma$ achievable iff $M\ge K_\gamma$; $B_{\min}=\lceil\log_2K_\gamma\rceil$; capacity-induced persistent failure; randomized deployment & witness and type arguments as in \citet{park2026regret}, including their union-control constant $|H|2^{|H|}$, which gives our $\eta_\Theta$; finite-memory testing \citep{hellman1970learning} as context & this paper (resource-indexed menus and frozen addressing)\\
Proposition~\ref{prop:oracle} (oracle-relative), Proposition~\ref{prop:monotone} (monotonicity and its failure) & --- & this paper (elementary)\\
Proposition~\ref{prop:barrier} (information barrier) & indistinguishability arguments; \citet{blackwell1953equivalent} & elementary; stated as a principle\\
Lemma~\ref{lem:capacity}(a),(b): capacity $\min(g,v+1)$ & all-negative path & this paper (elementary)\\
Lemma~\ref{lem:capacity}(c): capacity $2^{\min(g,v+1)}-1$ & search trees \citep{knuth1998art} & upper bound under generation and delivery constraints: this paper\\
Theorem~\ref{thm:continuous}, Propositions~\ref{prop:continuous-sharp}, \ref{prop:weak}: covering margin $d_*$ & Gaussian two-point testing; quantized estimation \citep{zhu2014quantized} as context & this paper (interval erosion; finite-sample bounds)\\
Corollary~\ref{cor:bridge}: binary measurements & Hoeffding; two-point bounds & bound transfer: this paper; exact binomial test classical\\
Theorem~\ref{thm:discrete}: covering order $t_*$ & covering designs \citep{gordon1995covering}; list testing \citep{asadi2023list}; exponent principle \citep{park2026regret} & reduction of budgets to $C(m,\ell,t)$, closed-form $K_\gamma$ and algorithm: this paper\\
Theorem~\ref{thm:conversion}: $D_{\cA}(F)\le2\tau\nu$ & increasing differences \citep{topkis1978minimizing}; monotone endpoint reasoning \citep{marquessilva2021explanations}; adaptive information \citep{novak1996power} & common worst corner under interleaved evidence and verification: this paper\\
Proposition~\ref{prop:certificates}, Corollary~\ref{cor:band} & oscillation certificates & this paper\\
Proposition~\ref{thm:planning}: planning NP-complete & CLIQUE \citep{karp1972reducibility}; dense subgraph \citep{feige2001dense} & direct consequence; no new hardness claimed\\
Theorem~\ref{thm:frontier}: exact frontier under adaptive interleaving & list testing; symmetry & this paper\\
Definition~\ref{def:sstar}, Proposition~\ref{prop:highorder}: $s_*$, high-order incompatibility & --- & this paper\\
Theorem~\ref{thm:sandwich}: sandwich with $\Omega_{s_*}$ & common-mass sandwich \citep[Sec.~V]{park2026regret}; two-point method \citep{lecam1973convergence} & identification via $s_*$, reduction to $\Omega_{s_*}$, constants: this paper\\
Lemma~\ref{lem:cyclic}: cyclic rotation & cycle lemma \citep{dvoretzky1947arrangements,spitzer1956combinatorial} & use for multi-world chains: this paper\\
Theorem~\ref{thm:second}: second order & lattice large deviations \citep{bahadur1960deviations}; the $e^{-nE}/\sqrt n$ prefactor is classical at fixed parameters & uniformity of the constants in $d$ and $\rho$, and $n_\delta$: this paper\\
Theorem~\ref{thm:compact}: compact codes & Helly number of Hamming balls \citep[Thm.~1.2]{alon2024helly}; sunflowers \citep{erdos1960intersection}; Sauer--Shelah & construction and bounds on $K_\gamma$: this paper\\
Proposition~\ref{prop:setcover}: computing $K_\gamma$ & greedy covering \citep{chvatal1979greedy}; inapproximability \citep{feige1998threshold} & direct consequence\\
Corollaries~\ref{cor:contK}, \ref{cor:faces}: explicit coverage numbers & --- & this paper (elementary)\\
\bottomrule
\end{tabular}
\caption{Attribution. ``This paper'' refers to the statement as formulated here. A literature search cannot establish global priority, and ``elementary'' results are listed for completeness.}
\label{tab:attribution}
\end{table}

\paragraph{Statistical decision theory and error exponents.} \citet{park2026regret} prove that, in finite Bayesian decision problems, the optimal regret exponent is the minimum multivariate Chernoff information over minimal incompatible state sets. Their proof also bounds the regret at every $n$ by common likelihood masses. Theorem~\ref{thm:coverage}(d) recovers this exponent in our worst-case formulation. What we add is the resource layer. The program menu is generated by online budgets. Frozen addressing restricts how many programs are usable. This yields the necessary and sufficient capacity $\lceil\log_2K_\gamma\rceil$, capacity-induced persistent failure, and explicit coverage numbers in structured models. Classical results cover binary Chernoff exponents \citep{chernoff1952measure}, Bayesian $M$-ary detection \citep{leang1997asymptotics}, hypothesis exclusion \citep{mishra2024antidistinguishability} and list testing \citep{asadi2023list}. The comparison of experiments \citep{blackwell1953equivalent} orders information structures uniformly over decision problems. We study the order induced by a single structured family of tasks, and this order depends on deployment budgets.

\paragraph{Information-based complexity and optimal recovery.} Information-based complexity \citep{traub1988information} and optimal recovery \citep{micchelli1977survey} measure how much information about an unknown input an algorithm needs. The power of adaptive information is surveyed by \citet{novak1996power}. Theorem~\ref{thm:conversion} is in this tradition. Its specific features are the following: a program learned in advance, coordinate evidence constrained by a family $\cA$, generated candidates, a radius verifier whose answers are themselves information, and a single delivered answer. The common worst corner identifies when adaptivity does not help, and Proposition~\ref{prop:certificates} gives a case where it does.

\paragraph{Rate distortion, quantized estimation, and task-dependent statistics.} Indirect rate distortion \citep{dobrushin1962information,witsenhausen1980indirect} and quantized estimation \citep{zhu2014quantized} trade bits for estimation error. The information bottleneck \citep{tishby2000information} and decision-focused learning \citep{elmachtoub2022smart} tailor learned representations or predictions to a downstream objective. Our capacity results differ in three ways. The loss is defined by executable programs under deployment budgets. Distortion is replaced by compatibility. The governing quantities, $K_\gamma$, $d_*$, $t_*$ and $s_*$, are covering numbers and margins, not mutual informations.

\paragraph{Memory and communication.} Learning with finite memory \citep{hellman1970learning} and memory--sample tradeoffs \citep{raz2018fast,steinhardt2016memory} constrain the state during learning. Communication complexity \citep{kushilevitz1997communication} constrains the bits exchanged between parties. Here the constrained object is the state handed from training to deployment, and its value is measured by the programs it can address.

\paragraph{Search, generation and verification.} Searching with comparisons, possibly with lies, is classical \citep{knuth1998art,pelc2002searching}. The constraint that only generated integers may be compared or delivered is what produces the capacities in Lemma~\ref{lem:capacity}. Generate-and-verify pipelines and their scaling are studied empirically and under reward and anti-concentration assumptions \citep{cobbe2021training,setlur2025scaling}. Our verifiers are exact truth ports with explicit budgets.

\paragraph{Compositional generalization.} \citet{wiedemer2023compositional} give support and structure conditions under which compositional generalization holds. In the end-to-end model the composition rule is likewise given, and the results quantify the finite-sample and resource side.

\section{Proofs for the general theory}\label{app:general}

\subsection{Multivariate Chernoff information}

\begin{lemma}[Minimax form]\label{lem:minimax}
Let $H$ be finite and let $P_\theta$, $\theta\in H$, be probability vectors with positive entries on a finite set $\cX$. Then
\[
C(H)=\min_{Q}\max_{\theta\in H}\KL(Q\|P_\theta)=-\log\min_{w\in\Delta_H}\sum_{x\in\cX}\prod_{\theta\in H}P_\theta(x)^{w_\theta}.
\]
The same holds for mutually absolutely continuous laws, after restricting $\cX$ to their common support.
\end{lemma}

\begin{proof}
For $w\in\Delta_H$ let $Z_w=\sum_x\prod_\theta P_\theta(x)^{w_\theta}>0$ and $Q_w=\prod_\theta P_\theta^{w_\theta}/Z_w$. For every distribution $Q$ on $\cX$, $\sum_\theta w_\theta\KL(Q\|P_\theta)=\KL(Q\|Q_w)-\log Z_w$, so $\min_Q\sum_\theta w_\theta\KL(Q\|P_\theta)=-\log Z_w$. The function $(Q,w)\mapsto\sum_\theta w_\theta\KL(Q\|P_\theta)$ is finite, continuous and convex in $Q$ on the compact simplex, and linear in $w$. By Sion's minimax theorem \citep{sion1958general},
\[
\min_Q\max_\theta\KL(Q\|P_\theta)=\min_Q\max_w\sum_\theta w_\theta\KL(Q\|P_\theta)=\max_w(-\log Z_w).
\]
\end{proof}

\begin{lemma}[Common mass and types]\label{lem:types}
Under the hypotheses of Lemma~\ref{lem:minimax}, for every $n$,
\[
(n+1)^{-|\cX|}\exp\Bigl(-n\max_{\theta\in H}\KL(Q_n\|P_\theta)\Bigr)\ \le\ \int\min_{\theta\in H}P_\theta^{\otimes n}\ \le\ (n+1)^{|\cX|}e^{-nC(H)}
\]
for every type $Q_n$ with denominator $n$, and consequently $-\frac1n\log\int\min_{\theta\in H}P^{\otimes n}_\theta\to C(H)$.
\end{lemma}

\begin{proof}
On a type class $T(Q)$ each $P_\theta^{\otimes n}$ is constant, equal to $\exp(-n[\mathrm H(Q)+\KL(Q\|P_\theta)])$, and
\[(n+1)^{-|\cX|}e^{n\mathrm H(Q)}\le|T(Q)|\le e^{n\mathrm H(Q)}\]
\citep{csiszar2011information}. Hence $\sum_{x\in T(Q)}\min_\theta P_\theta^{\otimes n}(x)=|T(Q)|e^{-n\mathrm H(Q)}e^{-n\max_\theta\KL(Q\|P_\theta)}$. The lower bound takes one class. The upper bound sums over at most $(n+1)^{|\cX|}$ classes and uses $\max_\theta\KL(Q\|P_\theta)\ge C(H)$. For the limit, choose types $Q_n\to Q^*$, where $Q^*$ attains $C(H)$; the divergences are continuous because the $P_\theta$ have full support.
\end{proof}

\subsection{Proof of the coverage theorem}\label{app:coverage}

\begin{proof}[Proof of Theorem~\ref{thm:coverage}]
\emph{(a)} Let $\pi_1,\dots,\pi_K$, $K=K_\gamma\le M$, be policies whose success regions cover every member of $\cH_\gamma$, and let $\phi(l)=\pi_l$; unused states map to $\pi_1$. Given $X_{1:n}$ with type $\hat Q_n$, let $H=\{\theta:\KL(\hat Q_n\|P_\theta)<\gamma\}$. If $H\ne\emptyset$, then $C(H)\le\max_{\theta\in H}\KL(\hat Q_n\|P_\theta)<\gamma$, so $H\in\cH_\gamma$ and the rule stores the least $l$ with $H\subseteq\Succ(\pi_l)$. If $H=\emptyset$, it stores $1$. If the true world satisfies $\KL(\hat Q_n\|P_\theta)<\gamma$, then $\theta\in H\subseteq\Succ(\pi_W)$. Hence
\[
F_n(\theta)\le\Prob_\theta\bigl(\KL(\hat Q_n\|P_\theta)\ge\gamma\bigr)=\sum_{Q:\,\KL(Q\|P_\theta)\ge\gamma}P_\theta^{\otimes n}(T(Q))\le(n+1)^{|\cX|}e^{-n\gamma},
\]
using $P_\theta^{\otimes n}(T(Q))\le e^{-n\KL(Q\|P_\theta)}$.

\emph{(b)} Fix $n$ and a scheme: a training kernel $k(w\mid x)$, an assignment and, for each state $w$, a distribution $\rho_w$ over $\Pi_\omega$ (a point mass for deterministic deployment). For a world $\theta$ put $f_w(\theta)=\rho_w(\{\pi:\theta\notin\Succ(\pi)\})$, the probability that state $w$ fails in $\theta$. We claim that some $H\in\cH_\gamma$ satisfies $\max_{\theta\in H}f_w(\theta)\ge\eta_\Theta$ for every $w$.

\emph{Deterministic deployment.} Each $\rho_w$ is a point mass at $\phi(w)$. If every $H\in\cH_\gamma$ lay in some $\Succ(\phi(w))$, the $M$ policies $\phi(1),\dots,\phi(M)$ would cover $\cH_\gamma$, contradicting $K_\gamma>M$. So some $H$ is missed by every state, and each $w$ has a world of $H$ with $f_w=1$.

\emph{Randomized deployment.} Let $\eta=(|\Theta|2^{|\Theta|})^{-1}$ and suppose the claim fails. Then every $H\in\cH_\gamma$ has a state $w(H)$ with $f_{w(H)}(\theta)<\eta$ for all $\theta\in H$. Let $\Pi_H$ be the set of policies that serve all of $H$. By a union bound over $H$, $\rho_{w(H)}(\Pi_H)>1-|H|\eta\ge1-2^{-|\Theta|}\ge1-1/|\cH_\gamma|$. For a fixed state $w$, the sets $H$ with $w(H)=w$ number at most $|\cH_\gamma|$, and each $\Pi_H$ has $\rho_w$-mass greater than $1-1/|\cH_\gamma|$. Hence their intersection has positive mass and contains a policy $\pi^*_w$ that serves all of them. The policies $\pi^*_1,\dots,\pi^*_M$ then cover $\cH_\gamma$, contradicting $K_\gamma>M$.

Given such an $H$, choose for each state $w$ a witness $\theta_w\in H$ with $f_w(\theta_w)\ge\eta_\Theta$. The worst case is at least the average over the $M$ witnesses, with repetitions allowed, so
\[
R_n\ge\frac1M\sum_{w}F_n(\theta_w)\ge\frac{\eta_\Theta}M\sum_x\sum_wP^{\otimes n}_{\theta_w}(x)k(w\mid x)\ge\frac{\eta_\Theta}M\sum_x\min_{\theta\in H}P^{\otimes n}_\theta(x),
\]
because $\sum_wk(w\mid x)=1$. Lemma~\ref{lem:types} gives the limit.

\emph{(c)} If $K_\gamma\le M$, (a) gives $\liminf-\frac1n\log R_n\ge\gamma$. If $K_\gamma>M$, then by (b) and Lemma~\ref{lem:types}, $\liminf-\frac1n\log R_n\le\max_{H\in\cH_\gamma}C(H)<\gamma$. Since $\Theta$ is finite, $\gamma\mapsto\cH_\gamma$ is a left-continuous step function: for $\gamma\uparrow\gamma^*$ the families $\cH_\gamma$ increase and stabilize at $\cH_{\gamma^*}$. Hence $K_{\gamma^*}=\lim_{\gamma\uparrow\gamma^*}K_\gamma$, so $\gamma^*=\sup\{\gamma:K_\gamma\le M\}$ satisfies $K_{\gamma^*}\le M$ and is achievable, while no larger exponent is. For vanishing risk, take $\gamma$ below the least positive value of $C$. Then $\cH_\gamma$ consists of the sets with $C(H)=0$, which are sets of worlds with identical laws, and $K_\gamma=K_{0+}$. If $M\ge K_{0+}$, (a) gives a positive exponent. If $M<K_{0+}$, (b) applies with $\int\min_{\theta\in H}P^{\otimes n}_\theta=1$ for every $n$. The statement about bits follows with $M=2^B$.

\emph{(d)} If every member of $\cH_\gamma$ is compatible, then $K_\gamma\le|\Pi_\omega|\le M$. Otherwise $\cH_\gamma$ contains an incompatible set, and therefore a minimally incompatible $G$ with $C(G)<\gamma$, so $K_\gamma=\infty$. Thus $K_\gamma\le M$ if and only if $\gamma\le\min_GC(G)$, and (c) gives $E^*(M)=\min_GC(G)$. If $\Theta$ is compatible, $K_\gamma=1$ for all $\gamma$.

\emph{(e)} If $\Theta\subseteq\Succ(\pi)$, store nothing and run $\pi$. Now let $H$ be an incompatible set of worlds with identical laws. Then $W$ has the same law under every $\theta\in H$. For each state $w$, $\sum_{\theta\in H}f_w(\theta)=\sum_\pi\rho_w(\pi)\,|H\setminus\Succ(\pi)|\ge1$, because no policy serves all of $H$. Averaging over $W$ gives $\sum_{\theta\in H}F_n(\theta)\ge1$, so some $\theta\in H$ has $F_n(\theta)\ge1/|H|$.
\end{proof}

\subsection{Oracle-relative targets, monotonicity, and the barrier}

\begin{proof}[Proof of Proposition~\ref{prop:oracle}]
Write $\Delta(\pi,\theta,Q)=\Vor(\theta,Q)-V(\pi;\theta,Q)\in\{0\}\cup[\Delta,1]$. For a world $\theta$ and the random policy $\pi_W$,
\[
\max_Q\E\,\Delta(\pi_W,\theta,Q)\ \ge\ \frac1{|\cQ|}\E\sum_Q\Delta(\pi_W,\theta,Q)\ \ge\ \frac{\Delta}{|\cQ|}\Prob(\theta\notin\Succ(\pi_W)),
\]
and $\max_Q\E\,\Delta(\pi_W,\theta,Q)\le\Prob(\theta\notin\Succ(\pi_W))$ because $\Delta\le1$ and $\Delta=0$ on $\{\theta\in\Succ(\pi_W)\}$. Taking the maximum over $\theta$ gives the claim. The constants do not depend on $n$, so exponents and capacity statements transfer. In (b) the failure probability is multiplied by $\Delta/|\cQ|$, and in (e) the persistent failure $1/|H|$ becomes $\Delta/(|\cQ|\,|H|)$.
\end{proof}

\begin{proof}[Proof of Proposition~\ref{prop:setcover}]
Success regions are closed under subsets in the sense that a region containing $H$ contains every subset of $H$, and every member of the finite family $\cH_\gamma$ lies in an inclusion-maximal member. Hence a set of policies covers $\cH_\gamma$ if and only if it covers $\cH_\gamma^{\max}$, and $K_\gamma$ is the set-cover number of the universe $\cH_\gamma^{\max}$ by the sets $\{H\in\cH_\gamma^{\max}:H\subseteq\Succ(\pi)\}$, $\pi\in\Pi_\omega$. The greedy guarantee is the classical bound of \citet{chvatal1979greedy}, $\sum_{i\le|U|}1/i\le1+\ln|U|$ times the optimum. For hardness, take a set-cover instance with universe $U$ and sets $S_1,\dots,S_r$. Put $\Theta=U$, assign distinct laws on a finite alphabet with full support, let $\gamma_0=\min_{\theta\ne\theta'}C(\{\theta,\theta'\})>0$, and let the menu consist of $r$ policies with $\Succ(\pi_j)=S_j$ (the target is part of the input). For $0<\gamma\le\gamma_0$ the family $\cH_\gamma$ consists of the singletons, so $K_\gamma\le M$ if and only if $U$ is covered by $M$ of the sets, which is NP-complete \citep{karp1972reducibility}.
\end{proof}

\begin{proof}[Proof of Proposition~\ref{prop:monotone}]
(i) Every policy that respects $\omega$ also respects $\omega'$, and its success region does not change when the target is fixed. Hence every face of $\cS_\omega$ is a face of $\cS_{\omega'}$, any cover of $\cH_\gamma$ from $\Pi_\omega$ is a cover from $\Pi_{\omega'}$, and $K_\gamma(\omega')\le K_\gamma(\omega)$. Part (c) of Theorem~\ref{thm:coverage} gives the exponent claim. (ii) Every admissible pair under $\omega$ is admissible under $\omega'$. (iii) Under an oracle-relative target, $\Vor(\cdot;\omega')\ge\Vor(\cdot;\omega)$, and a policy that was oracle-optimal under $\omega$ may no longer be. The numerical instance is Table~\ref{tab:example}: $(e,v)=(1,0)\to(1,1)$ raises $V^*_6$ from $0.2347$ to $0.6089$, while $E_*$ falls from $0.0438$ to $0.0217$.
\end{proof}

\begin{proof}[Proof of Proposition~\ref{prop:barrier}]
Let $\mathsf R$ denote the complete observation record and $A$ the delivered answer, which is a randomized function of $\mathsf R$ and of internal randomness independent of the world. If $\mathsf R$ has the same law under $\theta$ and $\theta'$, so does $A$, jointly with the task realization. For each realization the success sets $\mathcal T_\theta$ and $\mathcal T_{\theta'}$ of answers are disjoint, so $V(\theta)+V(\theta')=\E[\Ind\{A\in\mathcal T_\theta\}+\Ind\{A\in\mathcal T_{\theta'}\}]\le1$.
\end{proof}

\subsection{List capacity}\label{app:capacity}

\begin{proof}[Proof of Lemma~\ref{lem:capacity}]
\emph{(a) and (b).} For a deterministic controller, consider the path along which every verifier query is answered ``no''. It is determined by the controller alone. Let $u_1,\dots,u_r$ be the distinct candidates verified on this path, with $r\le v$, and let $u_0$ be the candidate it finally delivers. Since $u_0$ must have been generated, $r+1\le g$ if $u_0\notin\{u_i\}$, and $r\le g$ otherwise; in all cases the number of distinct values among $u_0,\dots,u_r$ is at most $\min(g,v+1)$, and there are none if $g=0$. If the target $t$ is such that every query on this path truly answers ``no'', the actual run follows the path and succeeds only if $u_0$ is acceptable. Otherwise the first ``yes'' occurs at some $u_i$ with $t$ acceptable for $u_i$. Hence the success set is contained in the set of targets acceptable for some $u_i$, $0\le i\le r$. For an equality verifier this set is $\{u_0,\dots,u_r\}$. For a radius verifier it is the union of the closed intervals $[u_i-\tau,u_i+\tau]$. Conversely, given $\nu$ values or intervals, the controller that verifies the first $\nu-1$ centers in turn, delivers on ``yes'', and otherwise delivers the last succeeds exactly on their union.

\emph{(c).} Consider a deterministic controller facing a comparator and an integer target. We first normalize it without changing its success set or increasing its budgets on any path. (1) A comparison whose answer is determined by earlier answers on the same path, including a repeated comparison, is removed and its answer simulated. (2) Each integer is regarded as generated at its first comparison or at its delivery, whichever comes first; integers that are generated but never compared or delivered on a path are dropped from that path. After normalization, every comparison on a path uses a fresh integer.

Let $F(g,v)$ be the largest success set of a normalized controller with at most $g$ charged generations and $v$ comparisons per path. Then $F(0,v)=0$, and $F(g,0)\le1$ for $g\ge1$. Let $g,v\ge1$. If the controller never compares, its success set has at most one element. Otherwise let $x$ be its first comparison. On ``$=$'' it can succeed only for $t=x$. On ``$<$'' it can succeed only for $t<x$; delivering or comparing $x$ again is useless there, and after removing such actions the continuation is a normalized controller with budgets $(g-1,v-1)$. The same holds for ``$>$''. Hence $F(g,v)\le\max\{1,1+2F(g-1,v-1)\}$. By induction on $\min(g,v+1)$, $F(g,v)\le2^{\min(g,v+1)}-1$. For achievability, a balanced binary search tree on at most $2^\nu-1$ integers has depth at most $\nu$. The controller generates the key at each node, compares at internal nodes, delivers on ``$=$'', and delivers the leaf key without comparison. This uses at most $\nu$ generations and $\nu-1$ comparisons per path.

\emph{Consequence.} A randomized controller is a mixture of deterministic ones. Each succeeds with probability $\mu(S)$, where $S$ is its success set, and the stated maxima over admissible $S$ are attained by the constructions above.
\end{proof}

\section{Proofs for the geometries of solvability}\label{app:geometry}

\subsection{Continuous worlds}\label{app:continuous}

\paragraph{Success supersets.} Fix a deterministic deployment controller. Simulate all radius-verifier answers as ``no''. The controller's frozen-bit probes then follow a decision tree with at most $2^c$ leaves; at most $2^B$ leaves are reachable, so at most $M$ leaves are reachable. By Lemma~\ref{lem:capacity}(b), every code that reaches leaf $j$ succeeds only on a set contained in
\[
A_j=[0,1]\cap\bigcup_{i=1}^{\Leq}[u_{j,i}-\tau,u_{j,i}+\tau],
\]
whose total length is at most $w$. We call $A_1,\dots,A_M$ the \emph{success supersets}. Replacing the actual success sets by these supersets only strengthens the controller, so lower bounds proved against the supersets are valid.

\begin{lemma}[Interval erosion]\label{lem:erosion}
Let $A_1,\dots,A_M\subseteq[0,1]$ be finite unions of closed intervals, each of total length at most $w<1$, and let $d_*<h<1$ with $Mw\ge1$. Then some closed interval of length $h$ in $[0,1]$ is contained in no $A_j$. If $Mw<1$, some point of $[0,1]$ lies in no $A_j$.
\end{lemma}

\begin{proof}
Write $A_j$ as a disjoint union of closed components of lengths $a_{j,1},a_{j,2},\dots$, so that $\sum_ia_{j,i}\le w$. The interval $[x,x+h]$ lies in $A_j$ only if it lies in one component. Hence the set of left endpoints $x\in[0,1-h]$ with $[x,x+h]\subseteq A_j$ has measure at most $\sum_i(a_{j,i}-h)_+\le(w-h)_+$. If $h\ge w$ this is zero. If $h<w$, then $M(w-h)<1-h$ is equivalent to $h(M-1)>Mw-1$, that is, to $h>d_*$. In either case the $M$ sets of good left endpoints have total measure less than $1-h$, so some left endpoint in $[0,1-h]$ is good for no $j$. The second claim holds because the total length is at most $Mw<1$.
\end{proof}

\begin{proof}[Proof of Theorem~\ref{thm:continuous}]
\emph{(i)} Cover $[0,1]$ by $\Leq$ closed intervals of length $2\tau$, verify the first $\Leq-1$ centers in turn, and deliver the last. This needs $g\ge\Leq$ and $v\ge\Leq-1$, both of which hold.

\emph{(ii)} By Lemma~\ref{lem:erosion}, some $\theta$ lies in no success superset, so it fails with probability one for every training rule.

\emph{(iii), lower bound.} Let $d_*<h<1$. By Lemma~\ref{lem:erosion} there is $x$ such that $[x,x+h]\not\subseteq A_j$ for every $j$. For each leaf $j$ choose $\theta_j\in[x,x+h]\setminus A_j$. Let $k_j(y)$ be the probability that the training rule, given data $y$, produces a code that reaches leaf $j$; then $\sum_jk_j(y)=1$. World $\theta_j$ fails whenever leaf $j$ is reached, so
\[
R_N\ge\frac1M\sum_j\Prob_{\theta_j}(\text{fail})\ge\frac1M\int\sum_jp_{\theta_j}^{(N)}(y)k_j(y)\,dy\ge\frac1M\int\min_jp^{(N)}_{\theta_j}.
\]
For the Gaussian location family, the sample mean $\bar Y\sim\mathcal N(\theta,\sigma^2/N)$ is sufficient, and the conditional law of the remaining coordinates does not depend on $\theta$. At each point the smallest of the densities is that of the parameter farthest from $\bar y$, which is $\min_j\theta_j$ or $\max_j\theta_j$. Hence $\int\min_jp^{(N)}_{\theta_j}=2\Phi(-\sqrt N\Delta/(2\sigma))$ with $\Delta=\max_j\theta_j-\min_j\theta_j\le h$. This is decreasing in $\Delta$, so $R_N\ge\frac2M\Phi(-\sqrt Nh/(2\sigma))$. Letting $h\downarrow d_*$ gives the claim.

\emph{(iii), upper bound.} Let $s=(1-w)/(M-1)$ and use the windows $[js,js+w]$, $j=0,\dots,M-1$. Adjacent windows overlap in an interval of length $w-s=d_*$. Training computes $\bar Y$, clips it to $[0,1]$, and selects the window $j$ whose assigned region $[m_{j-1},m_j)$ contains it, where $m_j=(j+\frac12)s+\frac w2$ is the midpoint of the $j$-th overlap (with $m_{-1}=-\infty$ and $m_{M-1}=\infty$). The distance from $m_j$ to the right end $js+w$ of window $j$ is $\frac w2-\frac s2=\frac{d_*}2$, and symmetrically on the left. Hence if $|\bar Y-\theta|\le d_*/2$, the selected window contains $\theta$; clipping does not increase the distance because $\theta\in[0,1]$. Deployment probes the $\lceil\log_2M\rceil\le\min(B,c)$ code bits, tiles the selected window by $\Leq$ intervals of length $2\tau$, verifies the first $\Leq-1$ centers, and delivers the last. So failure implies $|\bar Y-\theta|>d_*/2$, which has probability $2\Phi(-\sqrt Nd_*/(2\sigma))$.

\emph{Exponent and sample size.} By the Gaussian tail asymptotics $\log\Phi(-x)=-\frac{x^2}2(1+o(1))$, both bounds have exponent $d_*^2/(8\sigma^2)$. The constants $2/M$ and $2$ do not depend on $N$, so inverting gives $N_\delta=(8\sigma^2/d_*^2)\log(1/\delta)(1+o(1))$ at fixed $M$.
\end{proof}

\begin{proof}[Proof of Proposition~\ref{prop:continuous-sharp}]
\emph{(i)} Let $Mw=1$. If the supersets do not cover $[0,1]$, some world fails with probability one. Otherwise their total length is at most $Mw=1$ while their union has length one, so pairwise intersections are finite sets. Let $0=z_0<\dots<z_r=1$ be all component endpoints. Each open piece $(z_{i-1},z_i)$ is either contained in or disjoint from each $A_j$, and is contained in exactly one $A_j$, since two would overlap in positive length. Not all pieces belong to the same $A_j$, since $|A_j|\le w<1$. So some adjacent pieces $(z_{i-1},z_i)\subseteq A_a$ and $(z_i,z_{i+1})\subseteq A_b$ have $a\ne b$. For small $\eta>0$ let $\theta^-=z_i-\eta$ and $\theta^+=z_i+\eta$. Then $\theta^-$ lies only in $A_a$, and $\theta^+$ lies only in $A_b$, which is not $A_a$. Every leaf therefore fails at least one of the two worlds. Assigning each leaf to such a world gives $\Prob_{\theta^-}(\text{fail})+\Prob_{\theta^+}(\text{fail})\ge\int\min(p_{\theta^-}^{(N)},p_{\theta^+}^{(N)})=2\Phi(-\sqrt N\eta/\sigma)$. Hence $R_N\ge\Phi(-\sqrt N\eta/\sigma)\to\frac12$ as $\eta\downarrow0$.

\emph{(ii)} If $\Leq=1$, each $A_j$ is one interval. An interval that contains both $x$ and $x+h$ contains $[x,x+h]$, so each leaf misses one of the two endpoints. Assigning each leaf to a missed endpoint gives the two-point bound $R_N\ge\frac12\cdot2\Phi(-\sqrt Nh/(2\sigma))$, and we let $h\downarrow d_*$.

\emph{(iii)} With $M=2$ we have $d_*=2w-1$ and $\frac2M=1$, so the lower bound of Theorem~\ref{thm:continuous} is $\Phi(-\sqrt N(2w-1)/(2\sigma))$. For the upper bound use the windows $[0,w]$ and $[1-w,1]$ with threshold $\frac12$. On the overlap $[1-w,w]$ no error occurs. For $\theta<1-w$ the only error is $\bar Y\ge\frac12$, whose probability is increasing in $\theta$ and approaches $\Phi(-\sqrt N(2w-1)/(2\sigma))$ as $\theta\uparrow1-w$. The case $\theta>w$ is symmetric.
\end{proof}

\begin{proof}[Proof of Proposition~\ref{prop:weak}]
In the first contract, $\Leq=M-1$ and $\tau=\frac1{2M}$, so $w=1-\frac1M$. Split $[0,1]$ into $M$ segments of length $\frac1M$. State $j$ generates the centers of all segments except the $j$-th, verifies all but the last of them, and delivers the last. It succeeds exactly outside the open $j$-th segment. Training ignores the data and selects $j$ uniformly at random. A world fails only if the selected state is the one that misses it, which has probability at most $\frac1M$. In the second contract, $\Leq=1$, so Proposition~\ref{prop:continuous-sharp}(ii) gives $R_N\ge\Phi(-\sqrt Nd_*/(2\sigma))$, which tends to $\frac12$ as $\sigma\to\infty$. Both contracts have $M$ and $w$, and hence $d_*$ and the exponent, in common.
\end{proof}

\begin{proof}[Proof of Corollary~\ref{cor:bridge}]
\emph{Upper bound.} Estimate $\hat\theta=2(\bar Y-\frac14)$. Then $|\hat\theta-\theta|\le d_*/2$ if and only if $|\bar Y-p_\theta|\le d_*/4$. By Hoeffding's inequality the complement has probability at most $2e^{-Nd_*^2/8}$. Use the window decoder of Theorem~\ref{thm:continuous}.

\emph{Lower bound.} Take witnesses in $[x,x+h]$ as in the Gaussian case. For a fixed binary sequence, the likelihood $p^k(1-p)^{N-k}$ is log-concave in $p$, so over a parameter interval its minimum is attained at an endpoint. Hence $\min_jp^{(N)}_{\theta_j}\ge\min(f,g)$, where $f,g$ are the likelihoods at $x$ and $x+h$. By Cauchy--Schwarz and $\int\max(f,g)\le2$, $\int\min(f,g)\ge\frac12(\int\sqrt{fg})^2$. By Jensen's inequality, $(\int\sqrt{fg})^2\ge e^{-\KL(f\|g)}$. Per token, $\KL\le(p-p')^2/(p'(1-p'))\le\frac{16}3(h/2)^2=\frac43h^2$, because $p'\in[\frac14,\frac34]$. Hence $R_N\ge\frac1{2M}e^{-4Nh^2/3}$, and we let $h\downarrow d_*$.

\emph{Exact two-state value.} Let $M=2$, $\Leq=1$ and $\frac12\le w<1$. The two states must cover $[0,1]$, otherwise some world fails with probability one. Since $w<1$, the points $0$ and $1$ are served by different states, and enlarging the supersets to $[0,w]$ and $[1-w,1]$ only helps. For small $\epsilon$, the worlds $1-w-\epsilon$ and $w+\epsilon$ are each served by exactly one state. Their token probabilities are symmetric about $\frac12$, so the likelihood ratio is monotone in the count $J$ and equals one at $J=\frac N2$. The equal-prior two-point error therefore equals $\Prob(J>\frac N2)+\frac12\Prob(J=\frac N2)$, with $J\sim\Bin(N,p_-)$ and $p_-\to\frac12-\frac{d_*}4$. Majority voting with a fair coin at ties attains this value. For parity, compare a majority over $2r-1$ tokens with majority-with-coin over $2r$ tokens. The additional token changes the outcome only when the count is $r-1$ or $r$, and the two changes cancel because $\binom{2r-1}{r-1}=\binom{2r-1}{r}$.
\end{proof}

\begin{proof}[Proof of Corollary~\ref{cor:contK}]
By Theorem~\ref{thm:continuous} and Proposition~\ref{prop:continuous-sharp}, $M$ states achieve exponent $d_*^2/(8\sigma^2)$ when $Mw>1$ and exponent $0$ when $Mw\le1$, and no scheme does better. Exponent $\gamma$ is therefore achievable iff $Mw>1$ and $d_*\ge h_\gamma$. If $h_\gamma\ge w$ this fails for all $M$, since $d_*<w$. If $h_\gamma<w$, then $(Mw-1)/(M-1)\ge h_\gamma$ is equivalent to $M(w-h_\gamma)\ge1-h_\gamma$, and any such $M$ satisfies $Mw>1$ because $(1-h_\gamma)w>w-h_\gamma$ for $w<1$.
\end{proof}

\paragraph{Finite precision.} With rational $\tau$ and a public bound on $\sigma$, the window endpoints and candidates are rational. Truncate observations to $[-D,D]$ and round them to a grid of spacing $\eta$, where $D=1+\sigma\sqrt{2\log(2N/\alpha)}$. The upper bound then becomes $\alpha+2\Phi(-\sqrt N(d_*/2-\eta)/\sigma)$, and the leading exponent is preserved as $\eta\to0$ and $\alpha\to0$ exponentially. Training uses an integer accumulator of $O(\log N+\log(D/\eta))$ bits. Deployment uses $O(\Leq+\log M)$ word operations after probing the code, and $c$ counts only frozen-bit probes.

\subsection{Discrete worlds}\label{app:discrete}

\begin{proof}[Proof of Theorem~\ref{thm:discrete}]
If $\ell=m$, read the cell and verify its rules in turn; no training is needed.

\emph{(i) Lower bound.} Fix one cell and answer all evidence queries consistently with it. Under the all-``no'' simulation, each of the at most $M$ reachable leaves serves at most $\ell$ rules of the cell (Lemma~\ref{lem:capacity}(a)); enlarge each leaf's list to an $\ell$-set. Since $C(m,\ell,s)>M$ by maximality of $t_*$ (with $C(m,\ell,s)=\infty$ if $s>\ell$), some $s$-set $S$ of the cell is contained in no leaf's list. Assign each leaf to a rule of $S$ that it misses. Summing the failure probabilities over $S$ gives
\[
R_N\ge\frac1s\sum_{i\in S}\Prob_i(\text{fail})\ge\frac1s\sum_x\min_{i\in S}P_i^{\otimes N}(x).
\]
By Lemma~\ref{lem:minimax} and symmetry, $C(S)$ is attained at uniform weights:
\[
\min_w\sum_x\prod_{i\in S}P_i(x)^{w_i}=(K-s)q+q\min_w\sum_{j\in S}(p/q)^{w_j}=(K-s)q+s\,p^{1/s}q^{1-1/s},
\]
so $C(S)=E_s$. Lemma~\ref{lem:types} with $|\cX|=K$, and a type within $O(K/N)$ of the optimal $Q^*$, gives $-\frac1N\log R_N\le E_s+O(\frac{\log N}N)$. If $t_*=0$, then $s=1$ and the bound is $R_N\ge1$.

\emph{(ii) Upper bound.} Fix a public covering of all $t_*$-subsets of $[m]$ by $C(m,\ell,t_*)\le M$ blocks. Training counts the tokens $n_1,\dots,n_K$, selects the $t_*$ most frequent rules (ties broken publicly), takes their indices within their cells (at most $t_*$ distinct indices), and stores a block containing them, in $\lceil\log_2M\rceil\le\min(B,c)$ bits. Deployment reads the cell, generates the block's $\ell$ rules in that cell, verifies $\ell-1$ of them, and delivers the last. If the true rule $i$ is among the $t_*$ most frequent, it is served. Otherwise some set $J$ of $s-1$ other rules satisfies $n_j\ge n_i$ for all $j\in J$. Fix such a $J$ and let $r=p/q$. On this event $r^{(\sum_{j\in J}n_j-(s-1)n_i)/s}\ge1$, so by Markov's inequality the event has probability at most the expectation of this quantity. Documents are independent, and one document contributes the factor
\[
p\,r^{-(s-1)/s}+(s-1)q\,r^{1/s}+(K-s)q=s\,p^{1/s}q^{1-1/s}+(K-s)q=e^{-E_s}.
\]
A union bound over the $\binom{K-1}{s-1}$ choices of $J$ completes the proof.
\end{proof}

\paragraph{Costs.} Training uses $O(N+K\log K)$ word operations and $O(K\log(N+1))$ bits of workspace. The public covering table has at most $M\ell\lceil\log_2m\rceil$ bits and is world-independent. Constructing an optimal covering design can be expensive; any covering with at most $M$ blocks suffices for (ii), and published tables \citep{gordon1995covering} give many values.

\section{Proofs for resource conversion}\label{app:conversion}

\begin{lemma}[Fibers and the common worst corner]\label{lem:corner}
Let $F$ be continuous and coordinatewise nondecreasing on $[0,1]^p$. After reading $x_J=t$, the set of possible outputs is the interval $[F(t,\mathbf0),F(t,\mathbf1)]$, where the remaining coordinates are set to $\mathbf0$ and $\mathbf1$. If $F$ also has increasing differences, then for every $J$ the length $F(t,\mathbf1)-F(t,\mathbf0)$ is maximized at $t=\mathbf1$, with value $F(\mathbf1)-F(\mathbf1_J)$.
\end{lemma}

\begin{proof}
By monotonicity the extremes over the fiber are attained at the corners $\mathbf0$ and $\mathbf1$ of the unread coordinates. The fiber is connected and $F$ is continuous, so every intermediate value is attained. Increasing differences with $y=\mathbf0$, $z=\mathbf1$ on the unread coordinates and $s\le t$ on $J$ give $F(t,\mathbf1)-F(t,\mathbf0)\ge F(s,\mathbf1)-F(s,\mathbf0)$, so the length is maximal at $t=\mathbf1$. Coordinatewise increasing differences imply the blockwise version used here: raise the unread coordinates one at a time and the read ones one at a time, and add the resulting inequalities.
\end{proof}

\begin{proof}[Proof of Theorem~\ref{thm:conversion}]
\emph{Sufficiency.} Read a maximizer $J^*\in\cA$ of $F(\mathbf1_J)$. By Lemma~\ref{lem:corner} the remaining interval has length at most $D_{\cA}(F)\le2\tau\nu$. Split it into $\nu$ closed pieces of length at most $2\tau$. Verify the centers of the first $\nu-1$ pieces, delivering on ``yes'', and otherwise deliver the last center. This uses $\nu\le g$ generations and $\nu-1\le v$ verifications.

\emph{Necessity.} If $g=0$ nothing can be delivered. Otherwise fix any deterministic controller and follow the record in which every coordinate read returns $1$ and every verification returns ``no''. Let $J\in\cA$ be the set of coordinates read on this record, and let $u_0,\dots,u_r$ be the distinct candidates verified on it or delivered at its end. As in Lemma~\ref{lem:capacity}, there are at most $\nu$ of them. Every input $x$ with $x_J=\mathbf1$ such that $F(x)$ lies outside $\bigcup_i[u_i-\tau,u_i+\tau]$ is consistent with the record: its reads return $1$, its verifications truly answer ``no'', and the controller fails on it. By Lemma~\ref{lem:corner} such inputs realize every value in $[F(\mathbf1_J),F(\mathbf1)]$, an interval of length at least $D_{\cA}(F)$. If $D_{\cA}(F)>2\tau\nu$, the $\nu$ closed intervals cannot cover it, so some input fails. At equality, closed pieces of length $2\tau$ do cover the interval. If $\tau=0$, the condition reads $D_{\cA}(F)=0$.
\end{proof}

\begin{proof}[Proof of Proposition~\ref{prop:certificates}]
\emph{Lower certificate.} Fix $z$ attaining $\underline U$ up to an arbitrarily small slack, answer every read according to $z$, and answer every verification ``no''. The record reads some $J\in\cA$, and the fiber $\{x_J=z_J\}$ has an image interval of length $o(J,z)\ge\underline U$. If $\underline U>2\tau\nu$, the at most $\nu$ candidate intervals leave a value uncovered, and an input realizing it is consistent with the record.

\emph{Upper certificate.} Read a fixed $J$ that attains $\overline U$; every fiber image then has length at most $\overline U\le2\tau\nu$, and we tile it.

Under increasing differences, Lemma~\ref{lem:corner} shows that $o(J,z)\le o(J,\mathbf1)=F(\mathbf1)-F(\mathbf1_J)$ for every $z$. Hence $\overline U=\min_Jo(J,\mathbf1)$, and since $\underline U\ge\min_Jo(J,\mathbf1)$, we get $\underline U=\overline U$.

\emph{The example.} Let $F(t,x,y)=t+(1-t)x+ty$. Its partial derivatives are $1-x+y$, $1-t$ and $t$, all nonnegative, while $\partial_{tx}F=-1$. With two reads, $g=1$ and $v=0$, a controller can read $t$ first and then whichever of $x,y$ has the larger weight ($1-t$ or $t$). The remaining width is then $\min(t,1-t)\le\frac12$, and delivering the midpoint achieves error $\frac14$. Conversely, answering both reads with $\frac12$ leaves width at least $\frac12$: if $t$ is unread the width is $|1-x+y|=1$, and if $x$ or $y$ is unread it is $\frac12$. So the best adaptive worst-case width is $\frac12$. Each fixed pair leaves width at least $1$: with $\{t,x\}$ read, setting $t=1$ gives width $1$; with $\{t,y\}$ read, setting $t=0$ gives width $1$; with $\{x,y\}$ read, setting $x=0$ and $y=1$ gives width $2$. The formula $F(\mathbf1)-\max_{|J|\le2}F(\mathbf1_J)$ equals $2-2=0$, attained at $J=\{t,y\}$.
\end{proof}

\begin{proof}[Proof of Corollary~\ref{cor:band}]
\emph{Sufficiency.} Read the optimal $J$ for $\widehat F$. After the reads, the true conditional image lies in $[\widehat F(t,\mathbf0)-\eta,\widehat F(t,\mathbf1)+\eta]$, an interval of length at most $D_{\cA}(\widehat F)+2\eta$; tile it.

\emph{Impossibility.} For any controller, follow the record with all reads equal to $1$ and all verifications ``no''. On the fiber $x_J=\mathbf1$ the true $F$ takes the values $F(\mathbf1_J)\le\widehat F(\mathbf1_J)+\eta$ and $F(\mathbf1)\ge\widehat F(\mathbf1)-\eta$. Hence it realizes every value in an interval of length at least $\widehat F(\mathbf1)-\widehat F(\mathbf1_J)-2\eta\ge D_{\cA}(\widehat F)-2\eta>2\tau\nu$, by continuity. The rest of the argument is as in Theorem~\ref{thm:conversion}. This direction uses only the continuity of $F$.
\end{proof}

\begin{proof}[Proof of Corollary~\ref{cor:faces}]
(i) Read an optimal evidence set $J$ for $\widehat F$. For every $\theta\in H$ and every $x$, the conditional image of $F_\theta$ given $x_J$ lies in $[\widehat F(x_J,\mathbf0)-\eta,\widehat F(x_J,\mathbf1)+\eta]$, an interval of length at most $D_{\cA}(\widehat F)+2\eta\le2\tau\nu$, which does not depend on $\theta$. Tiling it by $\nu$ intervals of length $2\tau$ and executing Lemma~\ref{lem:capacity}(b) serves every $\theta\in H$ and every $x$ with one policy. (ii) By Theorem~\ref{thm:conversion} applied to $F_\theta$, no controller serves $\theta$ for every $x$, so no policy has $\theta$ in its success region. (iii) Each ball of (i) is contained in the success region of one policy, so $M'$ such policies cover $\cH_\gamma$ and $K_\gamma\le M'$; Theorem~\ref{thm:coverage}(a) applies to the menu formed by these policies.
\end{proof}

\begin{proof}[Proof of Proposition~\ref{thm:planning}]
$F_G$ is a polynomial with nonnegative coefficients, so it is continuous and monotone and its mixed partial derivatives are nonnegative. Evaluation sums over the edge list. Since $F_G(\mathbf1_J)=|E(G[J])|$, Theorem~\ref{thm:conversion} shows that a guarantee exists if and only if $g\ge1$ and $|E|-\max_{|J|\le e}|E(G[J])|\le2\tau\nu$. This is in NP: a set $J$ with $|J|\le e$ is a certificate, checkable in $O(p+|E|)$ time, and Theorem~\ref{thm:conversion} shows that non-adaptive reading suffices. For hardness, reduce from CLIQUE \citep{karp1972reducibility}. Given $(G,k)$, set $e=k$, $g=1$, $v=0$ (so $\nu=1$), and $2\tau=|E|-\binom k2$; if this quantity is negative, output a fixed no-instance. Since $G[J]$ has at most $\binom k2$ edges when $|J|\le k$, the condition $\max_{|J|\le k}|E(G[J])|\ge\binom k2$ holds if and only if $G$ has a $k$-clique.
\end{proof}

\paragraph{Randomized deployment after evidence.} Fix the read set and the values read, and let $[l,u]$ be the remaining output interval, with $K=\max(1,\lceil(u-l)/(2\tau)\rceil)$ and $\tau>0$. A deployment that may randomize has optimal worst-case success probability $\min(1,\nu/K)$ on this slice. For the upper bound, $K$ points with pairwise gaps larger than $2\tau$ fit in the interval; under a uniform prior on them, the all-``no'' path of any deterministic controller hits at most $\nu$ of them. For the lower bound, choose $\min(\nu,K)$ of $K$ equal pieces uniformly at random. This describes a conditional experiment. Extending it to randomized choice of the evidence itself is not covered by the theorem.

\section{Proofs for the end-to-end theorem}\label{app:endtoend}

\subsection{Compatibility of translated supports}

\begin{proof}[Proof of Corollary~\ref{cor:compat}]
Under count $j$ the residual target is uniform on $A_j$, so a controller with success set $S$ succeeds with probability $|S\cap A_j|/u$, and the oracle value is $\min(L,u)/u$. A single $S$ with $|S|\le L$ attains it for all $j\in\Gamma$ iff $|S\cap A_j|=\min(L,u)$ for all $j\in\Gamma$. If $L\le u$ this says $S\subseteq A_j$ with $|S|=L$, i.e.\ $S$ is an $L$-subset of $\bigcap_jA_j$. If $L\ge u$ it says $A_j\subseteq S$ for all $j$, i.e.\ $\bigcup_jA_j\subseteq S$, possible iff $|\bigcup_jA_j|\le L$. If $L=u$, $|\bigcap_jA_j|\ge u$ forces all $A_j$ to be equal, as does $|\bigcup_jA_j|\le u$. By Lemma~\ref{lem:capacity}(c) every such $S$ is executable; randomized controllers attain the oracle value for all $j$ only if almost every deterministic component does.
\end{proof}

\paragraph{Diameter and windows.} For a task $(I,m)$, $J(\theta)=\sum_{i\in I}\theta_i\oplus m_i$ changes by at most one when one coordinate of $\theta$ changes, so $|J(\theta)-J(\theta')|\le\dH(\theta_I,\theta'_I)\le\dH(\theta,\theta')$. Hence the counts realized by a set of diameter $r$ lie in a window of span $r$, and every set of worlds of diameter less than $s_*$ is compatible for every task (a subset of a compatible window is compatible, since both criteria in Corollary~\ref{cor:compat} are monotone under taking subsets).

\subsection{The exact frontier}\label{app:frontier}

\subsubsection{Reductions}

\paragraph{Reading the corpus.} A training rule that reads at most $n$ documents, possibly choosing adaptively which index to read next and when to stop, sees documents that are i.i.d.\ and independent of the unread ones. Coupling the $l$-th read document with $X_l$, its output is a measurable function of $(X_1,\dots,X_n)$ and its internal randomness $\xi$. We therefore assume $W=\mathcal A(X_{1:n},\xi)$.

\paragraph{Reading the prefixes first.} The external channel reveals bits of $Z_E$ in order of significance and at most $e$ of them; there are only $h$, so at most $e'$ are read on any path; likewise at most $c'$ local bits. Given any controller $\pi$, let $\pi'$ read $e'$ external and $c'$ local bits at the start and then run $\pi$, answering each of $\pi$'s read requests from the stored prefix. On every realization $\pi'$ performs the same generations and comparisons and delivers the same answer as $\pi$, so it succeeds with the same probability and respects all budgets. Write $Z_E=2^{h-e'}E_{\mathrm{hi}}+\alpha$ and $Z_C=2^{b-c'}C_{\mathrm{hi}}+\beta$. Then $Y=y_0+f(J)+z$ with known shift $y_0=2^{b+h-e'}E_{\mathrm{hi}}+2^{b-c'}C_{\mathrm{hi}}$ and residual $z=2^b\alpha+\beta$, uniform on $\cD$ in~\eqref{eq:residual} and independent of $(\theta,X,\xi,m,E_{\mathrm{hi}},C_{\mathrm{hi}})$ and of deployment randomness $\zeta$. Comparing $y$ with $Y$ is the same as comparing $y-y_0$ with $T=f(J)+z$.

\subsubsection{Proof of Theorem~\ref{thm:frontier}}

\emph{(i)} Knowing $\theta$ and the task, the controller knows $J$, so $T$ is uniform on $A_J$, a set of $u$ integers. By Lemma~\ref{lem:capacity}, the best success probability is $\min(L,u)/u$.

\emph{(iii) and the upper bound in (ii).} Fix $I$ and put the uniform prior on $\theta$. Let $\mathcal G$ be the $\sigma$-field generated by $X_{1:n}$, $\xi$, $m$, $E_{\mathrm{hi}}$, $C_{\mathrm{hi}}$ and $\zeta$. After the prefix reduction, the controller's success set $S$ (at most $L$ residual targets, by Lemma~\ref{lem:capacity}(c)) is $\mathcal G$-measurable. The variables $\xi,E_{\mathrm{hi}},C_{\mathrm{hi}},\zeta$ are independent of $(\theta,z)$, so the conditional law of $T$ given $\mathcal G$ is its conditional law given $(X_{1:n},m)$. Under the uniform prior the likelihood of $X_{1:n}$ factorizes over coordinates, and the factor for $\theta_i$ is proportional to $(1-\eps)^{n_i^{\theta_i}}\eps^{n_i^{1-\theta_i}}$, where $n_i^y$ counts tokens $y$ at address $i$; hence the $\theta_i$ are conditionally independent with $\Prob(\theta_i=1\mid X_{1:n})=\sigma(\lambda\mathsf s_i)$. So $\theta_i\oplus m_i$ are conditionally independent Bernoulli$(p_i)$, $J$ has the Poisson-binomial law $(q_j)$, and since $z$ is independent and uniform on $\cD$, $T$ has conditional law $\mu_n$ in~\eqref{eq:mu}. Therefore $\Prob(T\in S\mid\mathcal G)=\mu_n(S)\le\Top_L(\mu_n)$, and the Bayes success of every admissible pair is at most $\E\,\Top_L(\mu_n)$. The worst case over $\theta$ is at most the Bayes average, and this holds for each $I$.

\emph{Achievability and equality.} The posterior controller attains $\mu_n(S)=\Top_L(\mu_n)$ given $\mathcal G$ by Lemma~\ref{lem:capacity}(c), so its Bayes success equals $\E\,\Top_L(\mu_n)$. It remains to show that its success probability $\Psi(\theta,I)$ does not depend on $(\theta,I)$. Let $i\in[d]$ and let $\phi_i$ map a corpus to the corpus with all non-erased tokens at address $i$ flipped, and $m$ to $m\oplus e_i$ if $i\in I$ (and leave $m$ unchanged otherwise). Under $\theta$, $(X_{1:n},m)\sim P_\theta^{\otimes n}\otimes\mathrm{Unif}$; its image under $\phi_i$ has law $P_{\theta\oplus e_i}^{\otimes n}\otimes\mathrm{Unif}$ because the flip exchanges the token probabilities $1-\eps$ and $\eps$ at address $i$ and the polarity is uniform. The map sends $\mathsf s_i$ to $-\mathsf s_i$, hence $\sigma(\lambda\mathsf s_i)$ to $1-\sigma(\lambda\mathsf s_i)$; if $i\in I$ it also flips $m_i$, so $p_i$ is unchanged, and if $i\notin I$, $\mu_n$ does not involve $\mathsf s_i$. Thus $\mu_n$ is unchanged, and so is $J$, because $(\theta_i\oplus1)\oplus(m_i\oplus1)=\theta_i\oplus m_i$. With the same $(\xi,E_{\mathrm{hi}},C_{\mathrm{hi}},z,\zeta)$ the posterior controller (whose tie-breaking depends only on $\mu_n$) behaves identically, so $\Psi(\theta,I)=\Psi(\theta\oplus e_i,I)$; iterating, $\Psi$ is constant in $\theta$. For a permutation $\varpi$ of $[d]$, relabeling addresses maps $P^{\otimes n}_\theta$ to $P^{\otimes n}_{\theta\circ\varpi^{-1}}$ because addresses are uniform, and $\mu_n$ for $\varpi(I)$ computed on the relabeled corpus equals $\mu_n$ for $I$ on the original corpus, as $\prod_{i\in I}(1-p_i+p_iz)$ is symmetric. Hence $\Psi(\theta,I)=\Psi(\theta\circ\varpi^{-1},\varpi(I))$, and $\Psi$ is constant. A constant function equals its average, so the worst case of the posterior controller equals its Bayes value $\E\,\Top_L(\mu_n)$, which by (iii) is an upper bound for every pair. The counts $(\mathsf s_i)_{i\in[d]}$, each in $[-n,n]$, occupy $d\lceil\log_2(2n+1)\rceil$ bits. Alternatively the frozen state may store the corpus itself, one symbol of $[d]\times\{0,1\}\cup\{\text{erased}\}$ per document, in $n\lceil\log_2(2d+1)\rceil$ bits, from which deployment recomputes the counts on $I$; either suffices, and we prove no lower bound on the capacity needed for exactness.\hfill$\square$

\paragraph{Remarks.} If $g=0$ then $L=0$, $\Vor=0$ and $V^*_n=0$. If the polarity were fixed rather than uniform, the upper bound in (iii) would still hold for the Bayes value, but the worst case could be smaller.

\subsubsection{Gradient realization}\label{app:gradient}

Let the model have parameters $w\in\mathbb R^d$ and predict token $1$ at an informative position of address $i$ with probability $\pi_i(w)=\eps+(1-2\eps)\sigma(w_i)$; erased positions have a parameter-free prediction. The token loss is $\ell=-y\log\pi_i-(1-y)\log(1-\pi_i)$. At $w=0$, $\pi_i=\frac12$ and $\partial\pi_i/\partial w_i=(1-2\eps)\sigma'(0)=(1-2\eps)/4$, so $\partial\ell/\partial w_i=-2y\cdot\frac{1-2\eps}4+2(1-y)\frac{1-2\eps}4=(1-2\eps)(\tfrac12-y)$ and all other partial derivatives vanish. Summing over the corpus, $\partial_{w_i}\sum\ell=-(1-2\eps)\mathsf s_i/2$. One gradient step with step size $\eta=2\lambda/(1-2\eps)$ gives $w_i=\eta(1-2\eps)\mathsf s_i/2=\lambda\mathsf s_i$, the posterior log-odds; with the mean loss the step size is multiplied by $n$. For $n=0$ the update is empty and $w=0$. Because $w_i/\lambda=\mathsf s_i$ is an integer, the frozen state stores the integers $\mathsf s_i$ and a public scale, so no real number is used as a storage channel.

\subsubsection{Resources of the achieving program}

\begin{proposition}[Resources of the achieving program]\label{prop:resources}
Work in the word RAM with words of $w_0=\Theta(\log(n+d+2^{h+b}))$ bits, and let $\eps=\mathsf q_0/(\mathsf p_0+\mathsf q_0)$ be rational with $\mathsf p_0>\mathsf q_0$ coprime. Let $\ell_\star=\lceil n\log_2(\mathsf p_0+\mathsf q_0)\rceil+k+h+b+2$ and $\mathsf M(\ell)=O(\lceil\ell/w_0\rceil^2)$, the schoolbook cost of multiplying $\ell$-bit integers.
\begin{enumerate}
\item[(i)] \emph{Training} reads $n$ documents and uses $O(n+d)$ word operations and $O(d\log(n+1))$ bits of workspace. Its output, the counts $(\mathsf s_i)$, occupies the frozen state of $d\lceil\log_2(2n+1)\rceil$ bits. Workspace and frozen capacity are accounted separately.
\item[(ii)] \emph{Posterior computation.} For a task, computing the numerators $\mathsf p_0^{|\mathsf s_i|}$ and $\mathsf q_0^{|\mathsf s_i|}$ by repeated squaring takes $O(k\log(n+1))$ multiplications. The Poisson--binomial recursion over the $k$ addresses takes $O(k^2)$ multiplications and additions on integers of at most $\ell_\star$ bits, over the common denominator $\prod_{i\in I}(\mathsf p_0^{|\mathsf s_i|}+\mathsf q_0^{|\mathsf s_i|})\le(\mathsf p_0+\mathsf q_0)^n$. The total is $O((k\log(n+1)+k^2)\,\mathsf M(\ell_\star))$ word operations.
\item[(iii)] \emph{List selection.} Forming $u\mu_n$ performs $(k+1)u$ additions into a sorted array of at most $(k+1)u$ integer keys, each carrying an $\ell_\star$-bit mass; selecting the $L$ heaviest keys and sorting them by value takes $O((k+1)u\log((k+1)u+1))$ comparisons. The total is $O((k+1)u\log((k+1)u+1)\lceil\ell_\star/w_0\rceil)$ word operations and $O((k+1)u\,\ell_\star)$ bits of workspace.
\item[(iv)] \emph{Online budgets.} The deployment makes $e'$ evidence reads and $c'$ local probes, and on each path at most $\min(g,v+1)$ generations and $\min(g,v+1)-1$ comparisons.
\end{enumerate}
These are sufficient budgets for an exact implementation, not lower bounds. For $b=0$, $\cD$ is an interval and $\mu_n$ is piecewise constant on $O(k)$ intervals (at most $2k+2$ breakpoints), which removes the factor $u$. For $b>0$, $\mu_n$ has $O(k\,2^{h-e'})$ pieces, which reduces the factor $u$ to that number.
\end{proposition}

\begin{proof}
(i) Training is a single pass with one counter update per informative document. (ii) With $\mathsf s_i\ge0$ we have $\sigma(\lambda\mathsf s_i)=\mathsf p_0^{\mathsf s_i}/(\mathsf p_0^{\mathsf s_i}+\mathsf q_0^{\mathsf s_i})$; for $\mathsf s_i<0$ exchange $\mathsf p_0$ and $\mathsf q_0$, and polarity replaces $p_i$ by $1-p_i$. Repeated squaring computes a power with exponent at most $n$ in $O(\log(n+1))$ multiplications. Keeping all coefficients $q_j$ over the common denominator, every intermediate numerator is at most the denominator, so it has at most $n\log_2(\mathsf p_0+\mathsf q_0)+k$ bits. (iii) The masses $u\mu_n(y)=\sum_jq_j\Ind\{y-f(j)\in\cD\}$ are sums of at most $k+1$ such numerators. We generate the $(k+1)u$ pairs $(f(j)+z,q_j)$, sort them by key and merge equal keys. Selection compares masses that share a denominator, so each comparison is between integers. (iv) This follows from the comparison tree of Lemma~\ref{lem:capacity}(c).
\end{proof}

\subsection{Finite-sample sandwich}\label{app:sandwich}

\begin{lemma}[Common difficulty of a critical chain]\label{lem:chaindifficulty}
In the end-to-end model of Section~\ref{sec:endtoend}, let the worlds $\theta^0,\dots,\theta^s$ form a monotone flip chain on addresses $a_1,\dots,a_s$, and let $P_j$ be the one-document law under $\theta^j$. Then $C(\{P_0,\dots,P_s\})=C(\{P_0,P_s\})=-\log(1-\chi_s(1-a_\eps))$.
\end{lemma}

\begin{proof}
Let $\mathbb Q=\sqrt{P_0P_s}/\varphi_s$ with $\varphi_s=\sum\sqrt{P_0P_s}=1-\chi_s(1-a_\eps)$. Documents that are erased, or that hit an address outside $\{a_1,\dots,a_s\}$, have the same probability under all $P_j$, so $\mathbb Q/P_j=1/\varphi_s$ on them. At an address $a_l$, $\mathbb Q$ gives each token value probability $\rho\sqrt{\eps(1-\eps)}/(d\varphi_s)$, while $P_j$ gives the two values the probabilities $\rho\eps/d$ and $\rho(1-\eps)/d$ in some order. The contribution of the two tokens to $\KL(\mathbb Q\|P_j)$ is therefore
\[
\frac{\rho\sqrt{\eps(1-\eps)}}{d\varphi_s}\Bigl[\log\frac{\sqrt{\eps(1-\eps)}}{\varphi_s\eps}+\log\frac{\sqrt{\eps(1-\eps)}}{\varphi_s(1-\eps)}\Bigr]=\frac{2\rho\sqrt{\eps(1-\eps)}}{d\varphi_s}\log\frac1{\varphi_s},
\]
independent of the order. Summing gives $\KL(\mathbb Q\|P_j)=-\log\varphi_s$ for every $j$. Thus $C(\text{chain})\le-\log\varphi_s=C(\{P_0,P_s\})$, where the last equality uses the symmetry of the pair (optimal weight $\frac12$). Monotonicity of $C$ under inclusion gives the reverse inequality.
\end{proof}

\subsubsection{The two-point overlap}

\begin{lemma}\label{lem:omega}
Let $\theta,\theta'$ differ exactly on a set $F$ of $s$ addresses. (a) $\int\min(P^{n}_\theta,P^n_{\theta'})$ is given by~\eqref{eq:omega} and depends only on $s$ (and $n,\rho,\eps,d$). (b) $\Omega_{s'}(n)\le\Omega_s(n)$ for $s'\ge s$. (c) $\Omega_s(0)=1$ and $\Omega_s(n)\le\varphi_s^n$ with $\varphi_s=1-\chi_s(1-a_\eps)$.
\end{lemma}

\begin{proof}
(a) The likelihood ratio $P^n_\theta/P^n_{\theta'}$ of a corpus equals $\bigl(\frac{1-\eps}{\eps}\bigr)^{2\mathcal J-\mathcal M}$, where $\mathcal M$ is the number of informative documents with address in $F$ and $\mathcal J$ the number of those whose token agrees with $\theta$. Under $\theta$, $\mathcal M\sim\Bin(n,\chi_s)$ and $\mathcal J\mid\mathcal M\sim\Bin(\mathcal M,1-\eps)$; under $\theta'$ the roles of $\mathcal J$ and $\mathcal M-\mathcal J$ are exchanged. Hence $\int\min(P^n_\theta,P^n_{\theta'})=\Prob_\theta(\mathcal J<\mathcal M/2)+\Prob_\theta(\mathcal J=\mathcal M/2)+\Prob_{\theta'}(\mathcal J>\mathcal M/2)$ and the last term equals $\Prob_\theta(\mathcal J<\mathcal M/2)$. (b) The pair at distance $s'$ yields the statistic $(\mathcal M,\mathcal J)$ with rate $\chi_{s'}\ge\chi_s$. Independently erasing each counted document with probability $1-s/s'$ produces the statistic with rate $\chi_s$ under both hypotheses; since $(\mathcal M,\mathcal J)$ is sufficient and the total variation distance $1-\Omega$ does not increase under a common Markov kernel, $1-\Omega_s\le1-\Omega_{s'}$, that is, $\Omega_{s'}\le\Omega_s$. (c) $\min(p,q)\le\sqrt{pq}$ and the one-document Bhattacharyya coefficient is $1-\chi_s+\chi_sa_\eps$.
\end{proof}

\subsubsection{Lower bound}

\begin{proof}[Proof of the lower bound in Theorem~\ref{thm:sandwich} and Lemma~\ref{lem:cyclic}]
Let $s=s_*$ and let $\{t,\dots,t+s\}$ be an incompatible window. Fix $I$ and distinct $a_1,\dots,a_s\in I$ and $o_1,\dots,o_t\in I\setminus\{a_l\}$ (possible since $t+s\le k$). Let $\theta^j$ have ones exactly at $o_1,\dots,o_t,a_1,\dots,a_j$. Under $m=0$, $J(\theta^j)=t+j$.

Fix an admissible pair. Condition on the corpus $x=X_{1:n}$, on $m=0$ and on $(\xi,E_{\mathrm{hi}},C_{\mathrm{hi}},\zeta)$; the controller then has a deterministic success set $S$ of at most $L$ residual targets, and under $\theta^j$ it succeeds with conditional probability $|S\cap A_{t+j}|/u$. The deficits $\Delta_j=\min(L,u)-|S\cap A_{t+j}|$ are nonnegative integers and, by Corollary~\ref{cor:compat} and incompatibility of the window, not all zero; so $\sum_j\Delta_j\ge1$. Under other polarities deficits are also nonnegative. Writing $P_j=P^{\otimes n}_{\theta^j}$,
\begin{align*}
\max_j\bigl[\Vor-V(\theta^j)\bigr]&\ge\frac1{s+1}\sum_{j=0}^s\bigl[\Vor-V(\theta^j)\bigr]
\ge\frac{2^{-k}}{(s+1)u}\sum_x\sum_jP_j(x)\,\E[\Delta_j\mid x,m=0]\\
&\ge\frac{2^{-k}}{(s+1)u}\sum_x\min_jP_j(x).
\end{align*}
It remains to show $\sum_x\min_jP_j(x)\ge\Omega_s(n)/s$. For a corpus $x$ let $v_l=\lambda\mathsf s_{a_l}(x)$; flipping $a_l$ from $0$ to $1$ multiplies the likelihood by $e^{v_l}$, so $P_j(x)=P_0(x)e^{S_j}$ with $S_j=v_1+\dots+v_j$. For a rotation $\varrho\in\{0,\dots,s-1\}$, let $\theta^{\varrho,j}$ be the chain that flips $a_{\varrho+1},\dots,a_s,a_1,\dots$ in this cyclic order; its endpoints are $\theta^0$ and $\theta^s$ for every $\varrho$, and $P_{\theta^{\varrho,j}}(x)=P_0(x)e^{S^\varrho_j}$ with $S^\varrho_j=v_{\varrho+1}+\dots+v_{\varrho+j}$ (indices mod $s$).

\emph{Pointwise claim.} Let $\ell\in\{0,\dots,s-1\}$ minimize $S_\ell$ over $\{S_0,\dots,S_{s-1}\}$. Then $S^\ell_j\ge\min(0,S_s)$ for all $0\le j\le s$. Indeed for $j\le s-\ell-1$, $S^\ell_j=S_{\ell+j}-S_\ell\ge0$; for $j=s-\ell$, $S^\ell_j=S_s-S_\ell\ge S_s$ because $S_\ell\le S_0=0$; for $j>s-\ell$, $S^\ell_j=S_s-S_\ell+S_{j'}$ with $j'=j-s+\ell\in\{1,\dots,\ell\}\subseteq\{0,\dots,s-1\}$, and $S_{j'}\ge S_\ell$ by minimality, so $S^\ell_j\ge S_s$. Hence $\min_jP_{\theta^{\ell,j}}(x)=P_0(x)e^{\min(0,S_s)}=\min(P_0(x),P_s(x))$.

\emph{Symmetry.} Let $\varsigma$ be the cyclic permutation $a_l\mapsto a_{l+1}$ (indices mod $s$) of $[d]$, fixing all other addresses, acting on corpora by relabeling addresses. Since addresses are uniform and the noise is the same at every address, $\varsigma$ maps $P^{\otimes n}_\theta$ to $P^{\otimes n}_{\theta\circ\varsigma^{-1}}$. The endpoints are invariant ($\theta^0$ and $\theta^s$ are constant on $\{a_l\}$), and $\theta^{\varrho,j}\circ\varsigma^{-1}=\theta^{\varrho+1,j}$. Therefore $M_\varrho:=\sum_x\min_jP_{\theta^{\varrho,j}}(x)$ does not depend on $\varrho$, and
\begin{align*}
\sum_x\min_jP_j(x)=M_0=\frac1s\sum_{\varrho}M_\varrho&=\frac1s\sum_x\sum_\varrho\min_jP_{\theta^{\varrho,j}}(x)\\
&\ge\frac1s\sum_x\min(P_0(x),P_s(x))=\frac{\Omega_s(n)}s .
\end{align*}
The upper inequality $\sum_x\min_jP_j\le\sum_x\min(P_0,P_s)$ is immediate. Corpora of probability zero under all chain worlds do not contribute, and the argument covers $n=0$ and ties.
\end{proof}

The lower bound uses one task $I$; it therefore holds for the worst case over $I$ and for every training rule, including rules that retain the entire corpus. It does not require any pair in the chain to be incompatible.

\subsubsection{Upper bound}

\begin{proof}[Proof of the upper bound in Theorem~\ref{thm:sandwich}]
By Theorem~\ref{thm:frontier}, $R_n$ is the Bayes regret (uniform prior) of the posterior controller, which maximizes conditional success given $(X_{1:n},m)$ and the prefixes; hence $R_n$ is at most the Bayes regret of any controller using the same information. Fix $I$ and $m$, and the coordinates $\gamma=\theta_{[d]\setminus I}$. For a corpus $x$ and $\beta\in\{0,1\}^I$ let $\ell_\beta(x)=P^{\otimes n}_{(\beta,\gamma)}(x)$; the order of $\beta\mapsto\ell_\beta(x)$ does not depend on $\gamma$ because ratios depend only on $(\mathsf s_i)_{i\in I}$. Sort $\{0,1\}^I$ by decreasing $\ell_\beta(x)$ with a fixed tie-breaking rule and let $H(x)$ be the longest prefix of Hamming diameter at most $s-1$. Its counts under $m$ lie in a window of span at most $s-1<s_*$, so some list is oracle-optimal for all of $H(x)$; the auxiliary controller executes it. If $H(x)=\{0,1\}^I$ the regret is zero. Otherwise let $\alpha^+(x)$ be the first world after $H(x)$ and $\alpha^-(x)\in H(x)$ a world with $\dH(\alpha^-,\alpha^+)\ge s$. Every $\beta\notin H(x)$ appears no earlier than $\alpha^+(x)$, so $\ell_\beta(x)\le\min(\ell_{\alpha^-}(x),\ell_{\alpha^+}(x))\le\sum_{\{\alpha,\alpha'\}:\,\dH(\alpha,\alpha')\ge s}\min(\ell_\alpha(x),\ell_{\alpha'}(x))$. Since regret is at most one,
\[
\text{Bayes regret}\le2^{-k}\sum_x\sum_{\beta\notin H(x)}\ell_\beta(x)\le(1-2^{-k})\sum_{\{\alpha,\alpha'\}:\,\dH\ge s}\Omega_{\dH(\alpha,\alpha')}(n)\le\binom{2^k}2\Omega_s(n),
\]
using Lemma~\ref{lem:omega}(b). Average over $\gamma$ and $m$.
\end{proof}

\begin{proof}[Proof of the exponent]
By Lemma~\ref{lem:omega}(c), $\Omega_{s}(n)\le\varphi_s^n$, and Lemmas~\ref{lem:tilt}--\ref{lem:lattice} below (which do not use Theorem~\ref{thm:sandwich}) give $\Omega_s(n)\ge\frac12e^{-\lambda/2}\varphi_s^n/\sqrt{1+n}$; thus $-\frac1n\log\Omega_s(n)\to-\log\varphi_s=E_*$ for $s=s_*$, and~\eqref{eq:sandwich} transfers the limit to $R_n$.
\end{proof}

\subsection{Second-order learning complexity}\label{app:second}

Let $s=s_*$, $\chi =\chi _s$, $a=a_\eps$, $\varphi =1-(1-a)\chi =e^{-E_*}$ and $z=\chi a/\varphi $.

\begin{lemma}[Exact tilted representation]\label{lem:tilt}
$\Omega_s(n)=\varphi ^n\,\E\,e^{-\lambda|\Sigma_K|/2}$, where $K\sim\Bin(n,z)$ and, given $K$, $\Sigma_K$ is a sum of $K$ independent uniform signs.
\end{lemma}

\begin{proof}
For nonnegative $p,q$, $\min(p,q)=\sqrt{pq}\,e^{-|\log(p/q)|/2}$. Applying this to the corpus likelihoods, which are products,
\[
\min(P^{\otimes n}_\theta,P^{\otimes n}_{\theta'})=\prod_l\sqrt{P_\theta P_{\theta'}}(x_l)\cdot e^{-|\sum_l\psi(x_l)|},\qquad \psi=\tfrac12\log(P_\theta/P_{\theta'}).
\] Normalize the one-document geometric mean to the law $\tilde{\mathbb Q}=\sqrt{P_\theta P_{\theta'}}/\varphi $. Under $\tilde{\mathbb Q}$ a document is discriminative (informative, at a differing address) with probability $\chi \sqrt{\eps(1-\eps)}\cdot2/\varphi =z$, and then its token supports either world with probability $\frac12$, giving $\psi=\pm\lambda/2$; otherwise $\psi=0$.
\end{proof}

\begin{lemma}[Lattice bounds]\label{lem:lattice}
For every integer $\ell\ge0$, $\frac{e^{-\lambda/2}}{2\sqrt{\ell+1}}\le\E\,e^{-\lambda|\Sigma_\ell|/2}\le\frac{\sqrt2\coth(\lambda/4)}{\sqrt{\ell+1}}$.
\end{lemma}

\begin{proof}
The largest point mass of $\Sigma_\ell$ is $\beta_\ell=2^{-\ell}\binom{\ell}{\lfloor\ell/2\rfloor}$, attained at $0$ or $\pm1$. From $4^j/(2\sqrt j)\le\binom{2j}j\le4^j/\sqrt{\pi j}$ ($j\ge1$) and $\binom{2j+1}{j}=\frac12\binom{2j+2}{j+1}$ one gets $\frac1{2\sqrt{\ell+1}}\le\beta_\ell\le\sqrt{2/(\ell+1)}$ for all $\ell\ge0$. The lower bound follows from $\E e^{-\lambda|\Sigma_\ell|/2}\ge e^{-\lambda/2}\beta_\ell$. For the upper bound, $\Sigma_\ell$ takes values of one parity, spaced by $2$, so $\E e^{-\lambda|\Sigma_\ell|/2}\le\beta_\ell\sum_{y\equiv\ell\,(2)}e^{-\lambda|y|/2}\le\beta_\ell\bigl(1+2\sum_{j\ge1}e^{-\lambda j/2}\bigr)=\beta_\ell\coth(\lambda/4)$.
\end{proof}

\begin{proof}[Proof of Theorem~\ref{thm:second}]
\emph{Averaging over $K$.} Since $y\mapsto(1+y)^{-1/2}$ is convex, Jensen gives $\E(1+K)^{-1/2}\ge(1+nz)^{-1/2}$. The multiplicative Chernoff bound gives $\Prob(K<nz/2)\le e^{-nz/8}$, so
\[
\E(1+K)^{-1/2}\le e^{-nz/8}+(1+nz/2)^{-1/2}\le(1.4+\sqrt2)(1+nz)^{-1/2},
\]
using $\max_{y\ge0}e^{-y/8}\sqrt{1+y}=2e^{-3/8}<1.4$. With Lemmas~\ref{lem:tilt} and~\ref{lem:lattice},
\[
\frac{e^{-\lambda/2}}{2}\,\frac{\varphi^n}{\sqrt{1+nz}}\le\Omega_s(n)\le 3\sqrt2\coth(\lambda/4)\,\frac{\varphi^n}{\sqrt{1+nz}} .
\]
\emph{Comparing clocks.} Since $a\le \varphi \le1$, $a\chi\le z\le \chi $. Since $-\log(1-y)\in[y,y/(1-y)]$ and $1-(1-a)\chi =\varphi \ge a$, $(1-a)\chi \le E_*\le(1-a)\chi /a$. Therefore $\kappa_-(1+nE_*)\le1+nz\le\kappa_+(1+nE_*)$ with $\kappa_-=\min\{1,a^2/(1-a)\}$ and $\kappa_+=1/(1-a)$, which depend only on $\eps$. Combining with~\eqref{eq:sandwich}, $u\le2^{h+b}$ and $s_*\le k$ gives the two-sided bound with
\[
c_1=\frac{e^{-\lambda/2}\sqrt{1-a}}{2^{k+h+b+1}k(k+1)},\qquad C_1=\binom{2^k}2\frac{3\sqrt2\coth(\lambda/4)}{\sqrt{\kappa_-}} .
\]
\emph{Inversion.} $V^*_n$ is nondecreasing in $n$ (a rule can ignore documents), so $R_n$ is nonincreasing and $n_\delta=\min\{n:R_n\le\delta\}$. Put $\Phi(y)=y+\frac12\log(1+y)$, strictly increasing. If $\Phi(nE_*)\ge\log(C_1/\delta)$ then $R_n\le\delta$; if $R_n\le\delta$ then $\Phi(nE_*)\ge\log(c_1/\delta)$. Hence
\[
\Phi^{-1}(\log(c_1/\delta))/E_*\le n_\delta\le\Phi^{-1}(\log(C_1/\delta))/E_*+1 .
\]
For $\Lambda\ge2$, $y=\Phi^{-1}(\Lambda)$ satisfies $\Lambda-\frac12\log(1+\Lambda)\le y\le\Lambda$, so $\frac12\log(1+y)=\frac12\log\Lambda+O(1)$ and $y=\Lambda-\frac12\log\Lambda+O(1)$ with an absolute $O(1)$. With $\Lambda=\log(1/\delta)+O_{k,h,b,\eps}(1)$ we get $\log\Lambda=\log\log(1/\delta)+o(1)$, which yields the stated expansion for $\delta\le\delta_0(k,h,b,\eps)$. Since $V^*$ depends on $(N,T_{\mathrm{read}})$ only through $n=\min(N,T_{\mathrm{read}})$, both quantities must reach $n_\delta$.
\end{proof}

\paragraph{Uniformity and its limits.} The constants depend on $\eps$ through $\lambda$ and $a$. As $\eps\uparrow\frac12$, $\coth(\lambda/4)\to\infty$: the lattice is too fine for the square-root factor to appear until $K\gg\lambda^{-2}$. As $\eps\downarrow0$, $a\to0$ and $\kappa_-\to0$: in the noiseless limit $\Omega_s(n)=(1-\chi _s)^n$ has no square-root factor. For fixed $\eps$, the bound is uniform in $d$ and $\rho$, so with $d\to\infty$ the relevant clock is $nE_*\asymp n\rho s_*/d$, the expected number of discriminative documents.

\subsection{Compact frozen knowledge}\label{app:compact}

Throughout, $s=s_*<\infty$, $r=s-1$, and $B_r(y)$ is the Hamming ball of radius $r$ in $\{0,1\}^d$. Note $r\le k-1<d$.

\subsubsection{Compiler}

\begin{enumerate}
\item From the counts, let $\theta_0$ be the coordinatewise majority ($\theta_{0,i}=\Ind\{\mathsf s_i>0\}$). For $F\subseteq[d]$, the world $\theta_0\oplus\Ind_F$ has log-likelihood $\log P^{\otimes n}_{\theta_0}(x)-\lambda\,c(F)$, $c(F)=\sum_{i\in F}|\mathsf s_i|$. Order worlds by $(c(F),|F|,F)$ with a fixed lexicographic order on sets; this is a likelihood order. (The counts are $w_i/\lambda$ for the gradient learner of Appendix~\ref{app:gradient}.)
\item Let $H$ be the longest prefix of this order with diameter at most $r$. Since $\theta_0$ is first, $H\subseteq B_r(\theta_0)$. The first world outside $B_r(\theta_0)$ in the order is $\theta_0\oplus\Ind_{F^\circ}$ with $F^\circ$ the $r+1$ coordinates of smallest $|\mathsf s_i|$ (ties by index), because any $|F|\ge r+1$ has $c(F)\ge c(F^\circ)$ and $|F|\ge|F^\circ|$. Hence $H$ is the longest prefix of diameter at most $r$ of the order restricted to $B_r(\theta_0)\cup\{\theta_0\oplus\Ind_{F^\circ}\}$, a list of $M_r+1$ worlds, $M_r=\sum_{i\le r}\binom di$.
\item Extend $H$ greedily: scan $B_r(\theta_0)$ in a fixed order and add each world within distance $r$ of all current members. Call the result $\mathcal C$. Since compatibility with a set is inherited by subsets, one pass yields an inclusion-maximal set of diameter at most $r$ (a world rejected earlier remains rejected later).
\item By maximality and diameter, $\mathcal C=\bigcap_{y\in\mathcal C}B_r(y)$. Delete balls other than $B_r(\theta_0)$ one at a time as long as the intersection (computed inside $B_r(\theta_0)$) is unchanged. Let $\mathcal F$ be the final family. By \citet[Theorem~1.2]{alon2024helly} (valid since $d>r$) some $\mathcal F'\subseteq\mathcal F$ with $|\mathcal F'|\le2^{r+1}$ has the same intersection; any ball of $\mathcal F\setminus\mathcal F'$ other than $B_r(\theta_0)$ could have been deleted, so $|\mathcal F\setminus\{B_r(\theta_0)\}|\le2^{r+1}$.
\item Store $\theta_0$ ($d$ bits), the number of extra centers ($r+2$ bits) and each extra center as its difference set from $\theta_0$ (a size field of $\lceil\log_2(r+1)\rceil$ bits and at most $r$ addresses of $\lceil\log_2(d+1)\rceil$ bits). The total is $B_{\mathrm{upper}}$.
\end{enumerate}

\paragraph{Sunflower alternative.} The thinning step can also use a self-contained certificate. Write each extra center as a difference set of size $j\le r$. If, within one size layer, $r+2$ difference sets form a sunflower (pairwise intersections equal a common core), one of them can be deleted without changing the feasible set: for feasible $X$ ($|X|\le r$), the $r+1$ remaining petals are disjoint outside the core, so some remaining petal $F$ misses $X$ outside the core, and $|X\triangle F'|=|X|+j-2|X\cap F'|\le|X|+j-2|X\cap F|=|X\triangle F|\le r$ for the deleted $F'$ because $X\cap F'\supseteq X\cap\text{core}=X\cap F$. By the Erd\H{o}s--Rado sunflower lemma, a layer with more than $j!(r+1)^j$ sets contains such a sunflower. This gives at most $\sum_{j\le r}j!(r+1)^j$ extra centers, also $O_r(1)$.

\subsubsection{Decoder}

Let $R$ be the union of the stored difference sets. A world $\theta_0\oplus\Ind_X$ belongs to $\mathcal C$ iff $|X|\le r$ and $|X\triangle F|\le r$ for every stored $F$. Given a task $(I,m)$, if $X$ is feasible then so is $X'=X\cap(I\cup R)$: $|X'|\le|X|$, and for $F\subseteq R$, $|X'\triangle F|=|X'\setminus F|+|F\setminus X'|\le|X\setminus F|+|F\setminus X|$, using $F\setminus X'=F\setminus X$. Both give the same count $J$, which depends only on coordinates in $I$. So the counts realized by $\mathcal C$ are obtained by enumerating subsets of $I\cup R$ of size at most $r$, at most $\sum_{j\le r}\binom{k+|R|}{j}$ sets with $|R|\le r2^{r+1}$, each checked against $O_r(1)$ constraints and counted in $O(k)$ operations. They lie in a window of span at most $r<s_*$, so by Corollary~\ref{cor:compat} a common oracle-optimal list exists (for $L\le u$: the $L$ smallest elements of $\bigcap_jA_j$; for $L\ge u$: $\bigcup_jA_j$), executed by Lemma~\ref{lem:capacity}(c). Writing out the list costs a further $\Omega(\min(L,u))$ operations, which the bound above excludes. The decoder reads the $k$ bits of $\theta_0$ on $I$ and the certificate (random access to an address costs a word of $\lceil\log_2(d+1)\rceil$ bits).

\paragraph{Compilation cost.} Training reads the corpus once ($O(n+d)$ word operations). Sorting $M_r+1$ worlds costs $O(M_r\log(M_r+1))$ comparisons; prefix detection and greedy extension cost $O(rM_r^2)$; the deletion step computes intersections inside $B_r(\theta_0)$ in $O(rM_r^2)$ per test and at most $M_r$ tests, $O(rM_r^3)$ in total. For fixed $s_*$ this is polynomial in $d$ and independent of $n$. We do not claim these compilation costs are optimal.

\subsubsection{Proof of Theorem~\ref{thm:compact}}

\emph{(i)} If $\theta\in H$ then $\theta\in\mathcal C$ and the decoder attains $\Vor$ on every task. If $\theta\notin H$, let $\alpha^+$ be the first world after $H$ and $\alpha^-\in H$ with $\dH(\alpha^-,\alpha^+)\ge s$; as in Appendix~\ref{app:sandwich}, $P^{\otimes n}_\theta(x)\le\min(P^{\otimes n}_{\alpha^-}(x),P^{\otimes n}_{\alpha^+}(x))$. Summing over the $2^{d-1}\sum_{i\ge s}\binom di$ unordered pairs at distance at least $s$ and using Lemma~\ref{lem:omega}(b), $\Prob_\theta(\theta\notin H)\le2^{d-1}\sum_{i\ge s}\binom di\,\Omega_s(n)$, and the regret on this event is at most $\Vor$. The lower bound is that of Theorem~\ref{thm:sandwich}, valid for every scheme. Both sides are constant multiples of $\Omega_s(n)$ for fixed $d$, so Theorem~\ref{thm:second} applies with $d$-dependent constants.

\emph{(ii)} First let deployment be deterministic given the code $w$. Let $G_w$ be the set of worlds $\theta$ for which, for every task $(I,m)$ and every value of the read prefixes, the conditional success equals $\min(L,u)/u$. Suppose $G_w$ shatters a $k$-set $I$. Fix $m$ and a prefix value; the program is then a fixed controller with a success set $S$. The restrictions of $G_w$ to $I$ realize every pattern in $\{0,1\}^I$, hence every count in $\{0,\dots,k\}$, in particular every count of an incompatible window, and $S$ would be oracle-optimal for all of them, contradicting Corollary~\ref{cor:compat}. So $\mathrm{VC}(G_w)\le k-1$, and by the Sauer--Shelah lemma $|G_w|\le V_{d,k}$; thus $|\bigcup_wG_w|\le2^BV_{d,k}$. If $\theta\notin\bigcup_wG_w$, then whatever code training outputs, some task and prefix value have conditional deficit at least $1/u$; under uniform $(I,m)$ and uniform prefixes this triple has probability $\bigl(\binom dk2^k2^{e'+c'}\bigr)^{-1}$, so the regret averaged over tasks is at least $\bigl(\binom dk2^k2^{e'+c'}u\bigr)^{-1}=\mu_\circ$. With independent deployment randomness $\zeta$, apply this for each fixed $\zeta$ (the sets $G_{w,\zeta}$ satisfy the same bound) and average. For uniform $\theta$,
\[
\max_{\theta,I}R\ \ge\ \E_\theta\E_I R\ \ge\ \mu_\circ\bigl(1-2^{B-d}V_{d,k}\bigr),
\]
which gives the claim. If $R_n\to0$, then for large $n$ the regret is below $\mu_\circ(1-\eta)$ for every $\eta>0$, so $2^B\ge(1-\eta)2^d/V_{d,k}$ for all $\eta$, i.e.\ $B\ge d-\log_2V_{d,k}$.\hfill$\square$

For fixed $k$, $V_{d,k}\le(ed/(k-1))^{k-1}$, so $B_{\mathrm{lower}}\ge d-(k-1)\log_2(ed/(k-1))$, while $B_{\mathrm{upper}}=d+O_{s_*}(\log(d+1))$. The lower bound concerns vanishing regret; the gap parameter $\mu_\circ$ shrinks with $d$, so the lower bound does not address tasks with fixed positive error tolerance in growing dimension.


\section{Numerical verification}\label{app:numerics}

All scripts are in the \texttt{code/} directory of the source archive. They use exact rational arithmetic unless stated otherwise, use no random sampling at run time (one geometric check uses a recorded fixed seed), and require only Python with NumPy, SciPy and Matplotlib. The computations illustrate and check the theorems; they are not used in any proof.

\begin{itemize}
\item \texttt{coverage\_theorem\_check.py} (Theorem~\ref{thm:coverage}). The instance has four worlds, a three-letter alphabet, and a menu in which every pair of worlds, but no triple, is served. The script computes $C(H)$ for all subsets, the coverage numbers $K_\gamma$, and the predicted optimal exponent $E^*(M)$. It also computes the exact Bayes-optimal failure probability with the best menu of $M$ programs, for $n\in\{200,400,800\}$. Predictions: $E^*(1)=0$, $E^*(2)=0.0394$, $E^*(3)=E^*(4)=0.0712$. Empirical slopes of $-\log R_n$ between $n=400$ and $n=800$: $0$ (risk constant at $\frac12$), $0.0406$, $0.0720$ and $0.0719$. The residual gaps are consistent with polynomial prefactors.
\item \texttt{continuous\_geometry\_check.py} (Theorem~\ref{thm:continuous}). For $2{,}137$ random menus of $M$ unions of $\Leq$ intervals of length $2\tau$ with $w<1\le Mw$, the largest commonly coverable interval length never exceeds $d_*$; the largest observed difference is $-0.073$. The equally spaced windows attain $d_*$ to within $10^{-12}$. The window decoder's worst risk lies between the two bounds, and for $M=2$ it matches $\Phi(-\sqrt N(2w-1)/(2\sigma))$.
\item \texttt{unified\_check.py} (Theorems~\ref{thm:continuous} and~\ref{thm:discrete}). It finds the covering numbers $C(m,\ell,t)$ for $m\le5$, $\ell\le2$ by exhaustive search, and verifies by search over all public leaf lists that the largest coverable order equals $t_*$. It computes the exponents $E_2=0.4069$ and $E_3=0.5108$ for $K=3$, $p=0.8$, $q=0.1$, and the exact Bayes risks with one and two readable bits for $N\le96$. Between $N=36$ and $N=96$ the risks decay at slopes $0.415$ and $0.526$. A Gaussian grid check of the window decoder is also included.
\item \texttt{bern\_exact\_check.py} (Corollary~\ref{cor:bridge}). For four values of $w$ and $N\le16$, it verifies in exact rationals the two-state binomial formula, the parity plateaus $R_{2r}=R_{2r-1}$, and the step $R_{2r}-R_{2r+1}$. For $N\le7$ it enumerates all $2{,}032$ count-based decision rules to confirm Bayes optimality.
\item \texttt{check\_composition.py} (Theorems~\ref{thm:conversion} and~\ref{thm:planning}, Proposition~\ref{prop:certificates}). It checks the worst-corner fiber formula on a four-dimensional grid program for all read sets, and the sufficiency of the threshold for $108$ budget combinations by an exact decision program over interleaved reads and verifications. It confirms the adaptivity gap of the example in Proposition~\ref{prop:certificates}. It verifies the clique reduction on all $1{,}024$ graphs on five vertices and all thresholds.
\item \texttt{native\_frontier.py} (Theorem~\ref{thm:frontier}). For $d=5$, $k=4$, $\rho=\frac45$, $\eps=\frac14$, $h=3$, $g=3$, it checks in $26{,}576$ cases that the polynomial readout equals brute-force enumeration over the $16$ local worlds. It finds that the success under a fixed world equals the Bayes value for $n\le6$, and that $12$ exhaustive dynamic programs over all interleavings of evidence, comparisons and deliveries match $\Top_L$. It also checks the residual-geometry example and the high-order example (minimal incompatible family of size $4$, span $3$, tilted divergence $0.0664687$ to every chain law).
\item \texttt{exact\_regret.py}, \texttt{table\_rows.py}, \texttt{make\_fig.py} (Theorems~\ref{thm:frontier}--\ref{thm:second}, Table~\ref{tab:example}, Figure~\ref{fig:frontier}). They compute $V_n^*$ exactly for $n\le6$ and in floating point for $n\le100$. For $(e,v)=(1,0),(1,1),(0,1)$, the ratios $R_n/\Omega_{s_*}(n)$ lie in $[0.019,0.083]$, $[0.179,0.314]$ and $[0.010,0.028]$, inside the guaranteed intervals; the normalized regret $R_ne^{nE_*}\sqrt{1+nE_*}$ lies in $[0.017,0.049]$, $[0.170,0.190]$ and $[0.0086,0.017]$.
\item \texttt{second\_order.py} (Lemma~\ref{lem:tilt}). It verifies the exact tilted identity in rationals for $54$ parameter and sample-size combinations covering both parities.
\item \texttt{compact\_certificate.py} (Theorem~\ref{thm:compact}). Exhaustive checks: the prefix computed from the ball plus one witness ($2{,}504$ inputs), greedy maximality, intersection-preserving thinning, decoder lists on $208$ contracts, the projection to $I\cup R$ ($756$ checks), the pointwise rotation claim ($19{,}530$ arrays), exact chain-versus-pair overlaps for $n\le5$, and sizes of non-shattering families in $\{0,1\}^4$.
\end{itemize}

\end{document}